\RequirePackage{snapshot}
\documentclass[fleqn,usenatbib]{mnras}

\usepackage{newtxtext,newtxmath}
\usepackage[T1]{fontenc}

\DeclareRobustCommand{\VAN}[3]{#2}
\let\VANthebibliography\thebibliography
\def\thebibliography{\DeclareRobustCommand{\VAN}[3]{##3}\VANthebibliography}

\usepackage{graphicx}	% Including figure files
\usepackage{amsmath}	% Advanced maths commands
\usepackage{fontawesome} % GitHub icon
\usepackage{physunits}  % units
\usepackage{csvsimple}

\newcommand{\Msun}{M$_{\odot}$} % Msun (solar mass)
\newcommand{\vc}[1]{{\mathbf{#1}}}
\newcommand{\hatn}{\hat{\mathbf{n}}}
\newcommand{\quokka}{\textsc{Quokka}}

\newcommand{\aref}[1]{\hyperref[#1]{Appendix~\ref{#1}}}

\newcommand{\jcompphys}{J.~Comp.~Phys.}
\newcommand{\joss}{J.~Open Source Software}

\defcitealias{Colella_1984}{CW84}
\defcitealias{Skinner_2013}{SO13}

\usepackage{color}
\usepackage[normalem]{ulem}
\definecolor{darkgreen}{rgb}{0.13, 0.55, 0.13}

\title[Two-moment AMR radiation hydrodynamics on GPUs]{\textsc{Quokka}: A code for two-moment AMR radiation hydrodynamics on GPUs}

\author[B. D. Wibking et al.]{
    Benjamin D. Wibking$^{1,2}$\thanks{E-mail: ben.wibking@anu.edu.au (BDW)}
    and Mark R. Krumholz$^{1,2}$
\\
$^{1}$Research School of Astronomy \& Astrophysics, Mount Stromlo Observatory, Cotter Road, Weston Creek, ACT 2611 Australia\\
$^{2}$ARC Centre of Excellence for Astronomy in Three Dimensions (ASTRO-3D), Canberra ACT 2600, Australia
}

\date{Accepted XXX. Received YYY; in original form ZZZ}

\pubyear{2021}

\begin{document}
\label{firstpage}
\pagerange{\pageref{firstpage}--\pageref{lastpage}}
\maketitle

\begin{abstract}
    We present \quokka, a new subcycling-in-time, block-structured adaptive mesh refinement (AMR) radiation hydrodynamics code optimised for graphics processing units (GPUs). \quokka~solves the equations of hydrodynamics with the piecewise parabolic method (PPM) in a method-of-lines formulation, and handles radiative transfer via the variable Eddington tensor (VET) radiation moment equations with a local closure. We use the AMReX library to handle the adaptive mesh management. In order to maximise GPU performance, we combine explicit-in-time evolution of the radiation moment equations with the reduced speed-of-light approximation. We show results for a wide range of test problems for hydrodynamics, radiation, and coupled radiation hydrodynamics. On uniform grids in 3D on a single GPU, our code achieves  $>250$ million hydrodynamic updates per second and almost $40$ million radiation hydrodynamic updates per second. For radiation hydrodynamics problems on uniform grids in 3D, our code scales from 4 GPUs to 256 GPUs with an efficiency of 76 per cent. The code is publicly released under an open-source license on \faGithub\href{https://github.com/BenWibking/quokka-code}{GitHub}.
\end{abstract}


%%%%%%%%%%%%%%%%%%%%%%%%%%%%%%%%%%%%%%%%%%%%%%%%%%

%%%%%%%%%%%%%%%%% BODY OF PAPER %%%%%%%%%%%%%%%%%%

\section{Introduction}

%%% MRK version %%%

\subsection{Methods for radiation-hydrodynamics}

In many astrophysical systems, the radiation field carries a substantial portion of the total momentum and energy budget, and therefore must be included in any numerical simulation. However, simulating such systems, particularly at high resolution, presents a fundamental challenge in both physics and numerics. Part of this challenge is dimensionality: in full generality the radiation field is governed by the time-dependent equation of radiative transfer,
\begin{equation}
    \frac{1}{c} \frac{\partial}{\partial t} I_\nu + \hatn\cdot\nabla I_\nu = \eta_\nu - \kappa^{\rm (tot)}_\nu \rho I_\nu + \frac{1}{4\pi} \int \kappa^{\rm (sca)}_\nu \rho I_\nu \, d\Omega,
    \label{eq:transfer}
\end{equation}
where $I_\nu$ is the radiation intensity at frequency $\nu$ travelling in the direction specified by the unit vector $\hatn$, $\eta_\nu$ is the matter emissivity, $\rho$ is the matter density, and $\kappa_\nu^{\rm (tot)}$ and $\kappa_\nu^{\rm (sca)}$ are the total and scatting specific opacities, respectively. This is a time-dependent integro-differential equation with six dimensions: three positions, two angles (parameterised by $\hatn$), and the frequency. Full numerical solution of a problem of this dimensionality, at least if it must be done millions of times to run in tandem with a hydrodynamic (HD) or magnetohydrodynamic (MHD) simulation, remains out of reach for most applications.

Within the astrophysics community, there are two general approaches to numerical radiation hydrodynamics (RHD). One is characteristic methods, which solve \autoref{eq:transfer} (or its time-independent form), but only along rays pointing back to particular sources \citep[e.g.,][]{Abel02a, Rijkhorst06a, Krumholz07f} or rays randomly assigned by Monte Carlo \citep[e.g.,][]{Fleck71a, Tsang15a}. A second approach, which we will pursue here, is moment methods \citep[e.g.,][]{Mihalas_1984,Castor_2004}, whereby one takes moments of the transfer equation, thereby eliminating the angular dimensions of the problem. Fully eliminating the angular dependence requires some care, since in general for a moving medium the emissivity and opacity depend on direction, even if the medium itself interacts with light isotropically in its own rest frame. While one might guess velocity-dependent beaming effects are unimportant in non-relativistic problems, it turns out that one cannot formulate a consistent, energy-conserving theory of non-relativistic radiation-hydrodynamics without including them, at least to leading order \citep{Mihalas82a, Lowrie99a, Krumholz_2007}. Systems of moment equations where the radiation moments are written in the lab frame but the emissivity and absorption are written in the comoving frame, where they can be assumed to be isotropic, are known as mixed-frame formulations. This is the most common approach in modern RHD codes (although the comoving frame is increasingly popular; e.g., \citealt{Skinner_2019}). This results in a series of moment equations that one can solve in lieu of solving the equation of radiative transfer directly, but at the price of introducing the need for a closure relation to approximate the higher moments that appear in the equations being solved. Some authors also combine moment and characteristic methods \citep[e.g.,][]{Rosen17a}. While characteristic and moment methods are the only ones widely used in astrophysics, in terrestrial applications (for example neutron transport calculations for nuclear reactor design) there are two other widely-used approaches to handle the angular dependence of the transfer equation. One is to discretise the unit sphere using weighted quadratures (the discrete ordinates, or $S_n$, method; e.g., \citealt{Lathrop_1964,Adams97a}). The other is to expand the angular integration in spherical harmonics (the $P_n$ method; e.g., \citealt{Modest89a}). These methods can be much more computationally expensive than moment methods, possibly by several orders of magnitude.

One of the simplest closures is the flux-limited diffusion (FLD) approximation \citep{LeBlanc_1970,Alme_1973}, which retains only the first moment equation, for the radiation energy density, and closes the system by assuming both that the time derivative of the radiation flux is zero and that the Eddington tensor, defined as the ratio of the radiation pressure tensor to the radiation energy density, has a fixed value. However, a long-understood deficiency of flux-limited diffusion is that it cannot cast shadows (e.g., \citealt{Hayes_2003}), a property that can make a critical difference in the dynamics of some semitransparent problems (e.g., \citealt{Davis_2014}). A more accurate approximation is to evolve both the radiation energy density and the radiation flux, while still invoking a closure relation for the Eddington tensor; this is called a two-moment approach, since one solves for the first two moments of the radiation field. When the Eddington tensor is computed via a formal solution of the angle-dependent radiative transfer equation, we obtain the quasidiffusion or variable Eddington tensor (VET) method \citep{Goldin_1964}. When retaining a local closure for the radiation pressure tensor in terms of the radiation energy density $E$ and the flux $F$, we obtain a local VET method, commonly referred to as the M1 (`moment-one`) method \citep{Minerbo_1978,Levermore_1984,Dubroca_1999,Gonzalez_2007}. There are a number of moment-based astrophysical RHD codes, implementing a wide variety of closures, in wide use, including \textsc{Zeus} \citep{Turner_2001}, \textsc{FLASH} \citep{Fryxell_2000}, \textsc{Orion} \citep{Krumholz_2007, Shestakov08a, Li21a}, \textsc{Ramses} \citep{Commercon11a, Rosdahl_2013}, \textsc{Athena} \citep{Davis_2012, Jiang12a}, \textsc{Enzo} \citep{Reynolds_2009,Bryan_2014}, \textsc{Castro} \citep{Zhang_2011,Zhang_2013,Almgren_2020}, and \textsc{Fornax} \citep{Skinner_2019}, to give a partial list.

While the use of moment methods removes the dimensionality problem, it leaves a second problem, which is the strong mismatch in signal speeds between radiation and sound (or MHD) waves, which in a non-relativistic system travel at far less than the speed of light. This mismatch renders simple explicit methods, as are commonly used for HD and MHD, impractically slow for radiative transfer, due to the tiny time steps that would be imposed by the Courant-Friedrichs-Lewy (CFL) condition. For this reason, numerical methods for RHD either use an implicit method for the radiation part of the problem (e.g., \textsc{Zeus}, \textsc{Orion}, some versions of \textsc{Ramses} and \textsc{Athena}) or adopt the reduced speed of light approximation (\citealt{Gnedin_2001, Skinner_2013}; \textsc{Fornax}, other versions of \textsc{Ramses} and \textsc{Athena}). The reduced speed of light approximation (RSLA) consists of replacing the speed of light $c$ that defines the signal speed in the radiation moment equations with a lower speed $\hat{c}$, while keeping the terms that describe the rate of momentum and energy exchange between gas and radiation unchanged. The lower speed $\hat{c}$, while still substantially larger than the HD or MHD signal speeds, is close enough to those speeds to allow radiation time steps large enough to render explicit methods computationally feasible.

\subsection{Why a new radiation hydrodynamics code?}

In this paper we describe \quokka\footnote{Quadrilateral, Umbra-producing, Orthogonal, Kangaroo-conserving Kode for Astrophysics!}, a new code for RHD. In terms of the taxonomy introduced above, \quokka~is a two-moment code that uses the RSLA to allow an explicit treatment of radiation transport. \quokka~is also an adaptive mesh refinement (AMR) code, so it begins with a base grid at uniform spatial resolution, but then dynamically adds higher-resolution grids as needed to achieve user-specified accuracy goals \citep{Berger:1984, Berger:1989}. However, these features do not make \quokka~unique: \textsc{Orion} and \textsc{Ramses} (among others) offer moment-based AMR RMHD, while \textsc{Fornax} uses RSLA on a dendritic (though not adaptive) grid.

The unique feature of \quokka~is that it has been designed from the ground up to run efficiently on graphics processing units (GPUs). This design goal motivated our choice of both algorithms and low-level implementation details. While \quokka~is not the first GPU hydrodynamics code in astrophysics (others include \textsc{Gamer}, \citealt{Schive10a, Schive18a}, \textsc{Cholla}, \citealt{Schneider15a}, \textsc{Castro}, \citealt{Almgren_2020}, and \textsc{ARK-RT}, \citealt{Bloch_2021}), nor even the first AMR GPU code, it is the first to feature two-moment AMR RHD on GPUs.

Bringing RHD to GPUs creates some unique challenges. Contemporary compute nodes are often limited by data bandwidth, both in terms of moving data between main memory and the CPU or GPU, and in terms of moving data between CPUs or GPUs. For this reason, implicit methods generally have poor scalability, due to the need for global communications during an implicit solve (see, e.g., Appendix E of \citealt{Skinner_2019}). This imbalance between computation and communication is magnified on GPUs. Likewise, robust implicit methods require iterative sparse matrix solvers, which achieve lower peak efficiency on GPUs compared to CPUs due to their heavy use of indirect addressing and highly branching control flow. These considerations motivate our choice of an explicit RSLA method. They also motivate our choice of time integration strategy, which as we detail below has been designed to maximize computation (and therefore minimize the relative amount of communication) on each hydrodynamic timestep. We show that, with this strategy, we are able to achieve update computation rates of $>250$ million zone updates per second per GPU for pure HD, and nearly $40$ million for RHD. We also achieve $\ge 75\%$ parallel efficiency (compared to single-node performance) out to 256 GPUs. This combination of performance and scaling makes \quokka~substantially faster than any other public RHD code.

The remainder of this paper is organized as follows. In \autoref{section:methods} we introduce the set of equations that \quokka~solves, and detail our numerical methods for solving them. In \autoref{section:tests}, we present a wide range of tests that demonstrate the accuracy and capabilities of the code. \autoref{section:performance} covers our tests of code performance and scalability. In \autoref{section:discussion}, we discuss the range of applicability of our methods, and our plans for application and future code expansions. Finally, the code itself, including all test problems, is freely available on \faGithub\href{https://github.com/BenWibking/quokka-code}{GitHub} under an open-source license.

\section{Methods}
\label{section:methods}
\subsection{Equations}
We solve the equations of radiation hydrodynamics \citep{Pomraning_1973,Mihalas_1984,Castor_2004} for an inviscid, nonrelativistic fluid in local thermodynamic equilibrium in the mixed-frame formulation, where the radiation variables are defined in an inertial frame (i.e., Eulerian simulation coordinates) and the radiation-matter interaction terms are written in the frame comoving with the fluid, with the transformations between the frames accounted for via the addition of radiation-matter exchange terms that depend explicitly on the ratio of fluid velocity to the speed of light, $\beta=v/c$. In this first version of \quokka~we omit scattering, so that matter-radiation interaction is purely by emission and absorption. We write the equations as follows:
\begin{align}
    \frac{\partial \rho}{\partial t} + \nabla \cdot (\rho \vc{v})                              & = 0 \, ,       \\
    \frac{\partial (\rho \vc{v})}{\partial t} + \nabla \cdot (\rho \vc{v} \vc{v} + \mathsf{P}) & = \vc{G} \, ,  \\
    \frac{\partial E}{\partial t} + \nabla \cdot \left[(E + \mathsf{P})\vc{v}\right]           & = c G^0 \, ,   \\
    \frac{\partial E_r}{\partial t} + \nabla \cdot {\vc{F}_r}                                  & = -c G^0 \, ,  \\\
    \frac{1}{c^2}\frac{\partial \vc{F}_r}{\partial t} + \nabla \cdot \mathsf{P}_r              & = -\vc{G} \, ,
\end{align}
where $\rho$ is the gas density, $\vc{v}$ is the gas velocity, $E$ is the total energy density of the gas, $\mathsf{P} = \delta_{ij} P$ is the gas pressure tensor, $E_r$ is the radiation energy density, $F_r$ is the radiation flux, $\mathsf{P}_r$ is the radiation pressure tensor, $\nabla \cdot \rho \vc{v} \vc{v}$ denotes the sum $(\rho v_i v^j)_{,j}\,$, and $G^i$ is the radiation four-force, with $G^0$ the time-like component and $\vc{G}$ consisting of the space-like components. In the mixed-frame formulation, the radiation four-force to order $\beta$ is
\begin{align}
    -c G^0 = \rho (\kappa_P 4 \pi B - \kappa_E c E_r) + \rho \kappa_F \left( \frac{\vc{v}}{c} \cdot \vc{F}_r \right) \, , \\
    -\vc{G} = -\rho \kappa_F \frac{\vc{F}_r}{c} + \rho \kappa_P \left(\frac{4 \pi B}{c}\right) \frac{\vc{v}}{c} + \rho \kappa_F \frac{\vc{v}\mathsf{P}_r}{c} \, ,
\end{align}
where
$\kappa_F$, $\kappa_E$, and $\kappa_P$ are the flux-mean, energy-mean, and Planck-mean specific opacities evaluated in the comoving frame,
$B$ is the Planck function evaluated at the gas temperature,
and $\vc{v} \mathsf{P}_r$ is the tensor contraction $v_j \mathsf{P}_r^{ij}$ \citep{Mihalas_1984}. The latter two terms in the expression for $\vc{G}$ correspond to the relativistic work term of \cite{Krumholz_2007} and are only important in the regime $\beta \tau \gtrsim 1$ (where $\tau$ is a characteristic optical depth),
to which we cannot apply the RSLA (as discussed below),
so we neglect them. However, the term of order $\beta$ in the expression for $cG^0$ corresponds to the work done by the radiation force on the gas and can be the dominant term for problems of interest.

To apply the RSLA to these equations, we first rewrite the radiation moment equations so that they have a factor of exactly $1/c$ next to each of the time derivatives:
\begin{align}
    \frac{1}{c} \frac{\partial E_r}{\partial t} + \nabla \cdot \left( \frac{\vc{F}_r}{c} \right) = -G^0 \, , \\\
    \frac{1}{c} \frac{\partial}{\partial t} \left( \frac{\vc{F}_r}{c} \right) + \nabla \cdot \mathsf{P}_r = -\vc{G} \, ,
\end{align}
then we replace this $1/c$ factor with a factor of $1/\hat c$, where $\hat c$ is the reduced speed of light, and multiply through by factors of $\hat c$ to obtain the conservation law form of the reduced speed of light radiation moment equations (e.g., \citealt{Skinner_2013}):
\begin{align}
    \frac{\partial E_r}{\partial t} + \nabla \cdot \left( \frac{\hat c}{c} \vc{F}_r \right) = -\hat c G^0 \, , \\\
    \frac{\partial \vc{F_r}}{\partial t} + \nabla \cdot (c \hat c \, \mathsf{P}_r) = -c \hat c \, \vc{G} \, .
\end{align}
The maximum wave speed of this system of equations is bounded by $\hat c$ (as long as the flux satisfies causality, i.e. $F_r \leq cE_r$). As emphasised by \cite{Skinner_2013}, all other factors of $c$ remain unchanged, and, since the factors of $c$ are unchanged on the right-hand side of the hydrodynamic equations, the reduced speed of light radiation hydrodynamic system does not conserve total energy or momentum for $\hat{c} \neq c$.
When the left-hand side flux divergence terms are negligible, this nonconservation implies that the equilibrium temperature of the reduced speed of light system is slightly modified with respect to the correct equilibrium temperature, implying that we cannot apply the RSLA to problems in the equilibrium diffusion limit in general (see section \ref{section:equilibrium}).

Writing out the right-hand side terms explicitly, we obtain
\begin{align}
    \label{eq:hydro_continuity}
    \frac{\partial \rho}{\partial t} + \nabla \cdot (\rho \vc{v})                              & = 0 \, ,                                                                                                                                         \\
    \label{eq:hydro_momentum}
    \frac{\partial (\rho \vc{v})}{\partial t} + \nabla \cdot (\rho \vc{v} \vc{v} + \mathsf{P}) & = \rho \kappa_F {\vc{F}_r / c} \, ,                                                                                                              \\
    \label{eq:hydro_energy}
    \frac{\partial E}{\partial t} + \nabla \cdot \left[(E + \mathsf{P})\vc{v}\right]           & = -c \rho (\kappa_P a_r T^4 - \kappa_E E_r) - \rho \kappa_F \left( \frac{\vc{v}}{c} \cdot \vc{F}_r \right) \, ,                                  \\
    \label{eq:rad_energy}
    \frac{\partial E_r}{\partial t} + \nabla \cdot \left( \frac{\hat c}{c} \vc{F}_r \right)    & = \hat c \rho \left(\kappa_P a_r T^4 - \kappa_E E_r \right) + \rho \kappa_F \left( \frac{\hat c}{c} \frac{\vc{v}}{c} \cdot \vc{F}_r \right) \, , \\\
    \label{eq:rad_flux}
    \frac{\partial \vc{F}_r}{\partial t} + \nabla \cdot (c \hat c \, \mathsf{P}_r)             & = -\hat c \rho \kappa_F \vc{F}_r \, .
\end{align}
These equations make no approximations about the frequency dependence of the radiation field. However, for computational tractability, in what follows we will approximate $\kappa_F$ with the Rosseland mean opacity $\kappa_R$, which yields the correct radiation force in the diffusion limit, and approximate $\kappa_E$ with the Planck mean opacity $\kappa_P$, which yields the correct energy absorption and emission in the optically-thin limit for fluids at rest \citep{Mihalas_1984}. However, we emphasise that the choice to set $\kappa_F \approx \kappa_R$ and $\kappa_E \approx \kappa_P$ is an additional approximation, and that others might be preferable depending on the physical system being simulated.  In future work, we plan to address the limitations of these approximate grey opacities via an extension of our method to the multigroup solution of the radiation moment equations.  Our present set of equations is sufficient for grey nonrelativistic radiation hydrodynamics in the semi-transparent regime, where we can neglect the `relativistic work term' that is important only in the dynamic diffusion ($\beta \tau \gtrsim 1$) regime, as described earlier.

\subsection{Solution method}

We solve the system formed by \autoref{eq:hydro_continuity}--\autoref{eq:rad_flux} using an operator split approach, whereby we first advance the hydrodynamic transport subsystem (\autoref{sssec:hydro}), then the radiation transport subsystem (\autoref{sssec:radiation}), and finally update the local coupling terms (\autoref{sssec:coupling}). The first subsystem uses a single explicit update step, the second a set of subcycled explicit updates, and the third a purely local implicit update. We describe each of these steps below.

This update cycle operates within a \citet{Berger:1984} / \citet{Berger:1989} adaptive mesh refinement (AMR) framework, whereby each spatial variable is represented by a volume average in each cell, on a rectangular, Cartesian grid. We cover the entire computational domain with a coarse grid with cell spacings $\Delta x_0$, $\Delta y_0$, $\Delta z_0$ in the $x$, $y$, and $z$ directions; the grid spacings need not be the same, but for most applications we choose them to be the same. We denote this coarse grid level 0. We then dynamically add (or remove) additional, finer grids over parts of the domain in response to user-specified refinement criteria. We denote these additional levels 1, 2, $\ldots$, with each grid on level $l$ having cells a factor of $2$ smaller than those on level $l-1$, so that the cell spacing on level $l$ is $\Delta x_0/2^l$ in the $x$ direction, and similarly for $y$ and $z$. We use only factor of 2 refinements in order to minimize numerical glitches arising from the discontinuous change in resolution, which can arise especially in problems where shocks cross the coarse-fine mesh interface at an oblique angle (e.g., \citealt{Fryxell_2000}). When adding finer grids we conservatively interpolate the underlying coarse data (using linear interpolation for robustness, regardless of the spatial reconstruction used to compute the fluxes), and when removing finer grids we conservatively average down the fine data. Time steps on different AMR levels are sub-cycled, such that the time step on level $l$ is $\Delta t_l = \Delta t_0/2^l$. At the end of every two time steps on level $l > 0$, we perform a synchronization step to ensure that we maintain machine-precision conservation for all conserved quantities (\autoref{sssec:sync}).

Our implementation of AMR in \quokka~uses the lower-level \texttt{AMRCore} interface provided by the \textsc{AMReX} library \citep{AMReX_JOSS, the_amrex_development_team_2021_5363443} for adaptive mesh grid generation and coarse/fine grid interpolation, domain decomposition, and parallel communication. In addition to solving the radiation hydrodynamics equations, \textsc{Quokka} itself handles the timestepping and mesh refinement criteria.

\subsubsection{Hydrodynamics}
\label{sssec:hydro}
For the solution of the hydrodynamic subsystem (\autoref{eq:hydro_continuity}--\autoref{eq:hydro_energy}, omitting the matter-radiation coupling terms on the right hand sides), we adopt a method-of-lines (or semi-discrete) approach, discretizing the spatial variables while initially keeping the time variable continuous, thereby transforming the partial differential equations into a large set of ordinary differential equations that can be integrated in time using a standard ordinary differential equation (ODE) integrator \citep{Hyman_1979,Jameson_1981}. For the latter, we use the second-order strong stability preserving Runge-Kutta method (RK2-SSP; \citealt{Shu_1988}). Such an approach has been successfully employed in several recent astrophysical hydrodynamics codes \citep{Skinner_2019,Stone_2020}.

We schematically write the timestep $\Delta t$ used for the RK2-SSP integration as
\begin{align}
    \Delta t = C_0 \, \frac{\Delta x}{|\lambda|} \, ,
\end{align}
where $\Delta x$ is the minimum grid spacing, $|\lambda|$ is a maximum signal speed, and $C_0$ is a stability coefficient.
Analysis of the stability polynomial of a Runge-Kutta integrator applied to the linearized hydrodynamics equations \citep{Colella_2011,McCorquodale_2011} yields a value for $|\lambda|$ of
\begin{align}
    |\lambda| = \max \sum_{d=1}^{D} [(\vc{v} \cdot \vc{\hat{e}_d}) + c_s] \, ,
    \label{eq:colella_stability}
\end{align}
where $\vc{v}$ is the fluid velocity, $\vc{\hat{e}}$ is the unit vector in coordinate direction $d$, $c_s$ is the adiabatic sound speed, and the maximum is taken over all cells. However, even in the linear case, such an eigenvalue analysis gives, in general, only a necessary condition for stability and not a sufficient condition \citep{Reddy_1992}. For this reason, we more conservatively estimate the value of $|\lambda|$ as
\begin{align}
    |\lambda| = D \max (|\vc{v}| + c_s) \, ,
    \label{eq:signal}
\end{align}
and compute the timestep $\Delta t$ on each AMR level as
\begin{align}
    \Delta t = \frac{C_0}{D} \frac{\Delta x}{\max (|\vc{v}| + c_s)} \, ,
\end{align}
where we define the dimensionless factor $C_0 / D$ to be the CFL number so as to be consistent with its standard definition in one spatial dimension.\footnote{This is the same timestep criterion used in the \textsc{Fornax} code, with $D=3$ and $\Delta x = \Delta r / \sqrt{2}$; see Eq. 37 of \citealt{Skinner_2019}.} When written in this form, for both forward Euler and RK2-SSP, the maximum stable coefficient $C_0$ for a system of constant-coefficient, linear equations is $C_0 = 1$. The maximum stable CFL number in 3D for the RK2-SSP integrator is therefore $1/3$.
We note that it is not sufficient to estimate $|\lambda|$ as $\max (|\vc{v}| + c_s)$, since the component-wise sum of the velocities may exceed the vector magnitude $|\vc{v}|$ and therefore violate the lower bound given by \autoref{eq:colella_stability}.\footnote{As an example, consider the velocity vector with unit magnitude $|\vc{v}|$ and equal components $v_x = v_y = v_z$. Then each component $v_i = \sqrt{3}/3$ and the sum of components $\sum_{d=1}^{3} \vc{v} \cdot \vc{\hat{e}_d} = \sqrt{3} \approx 1.732$.}

We find that such a method-of-lines scheme is not stable when combining higher-order spatial reconstruction with forward Euler time integration. However, we find it is stable for timesteps satisfying the above timestep criterion when used with higher-order (second-order or higher) Runge-Kutta methods. We note that this stability problem with forward Euler is also found by \cite{Stone_2020} in the method-of-lines implementation of \textsc{Athena++}, but does not appear for single-step integrators that average in time the reconstructed profiles of characteristic waves over the cell interfaces, as done in the original version of PPM \citep{Colella_1984}.

As \cite{Skinner_2019} notes, in contrast to fully-discrete unsplit hydrodynamic methods such as the corner transport upwind (CTU) method \citep{Colella_1990}, the coupling across corners of each cell is achieved via the use of a multi-stage time integrator, rather than via direct computation of fluxes from diagonal neighbors of each cell. While we are formally limited to a smaller timestep compared to the CTU method (due to the factor of $1/D$), our method may be more robust in practice, as the CTU integrator has been found to be unstable in supersonic turbulence with strong radiative cooling unless very small ($\lesssim 0.1$) CFL numbers are employed \citep{Schneider_2017}.

We reconstruct the hydrodynamic variables on each face of each cell from the cell-average variables of the neighbouring cells. We perform this reconstruction using the piecewise parabolic method \citep[PPM;][hereafter \citetalias{Colella_1984}]{Colella_1984} using the primitive hydrodynamic variables (density, velocity, and pressure). As is standard, the conversion from conserved (density, momentum, and energy) to primitive variables is carried out assuming that the volume average and cell centered states are equivalent, which is an approximation accurate to $\mathcal{O}(\Delta x^2)$. As noted by several authors, the PPM algorithm is therefore formally second-order accurate in spatial resolution.\footnote{ We note that there exist fully fourth-order versions of PPM \citep{Felker_2018}, but because fourth-order accuracy does not permit local source terms to be evaluated independently for each cell, we choose to implement a second-order method.} After the primitive variables have been defined, for the reconstruction step proper, we use the standard interface-centered PPM stencil:
\begin{align}
    q_{j+1/2} = \frac{7}{12} (q_j + q_{j+1}) - \frac{1}{12} (q_{j+2} + q_{j-1}).
\end{align}
We follow the implementation of \cite{Stone_2020} in re-grouping the above terms symmetrically with respect to the interface ${i+{1/2}}$ so as to preserve exact symmetry in floating point arithmetic.

We do not perform the slope-limiting and contact steepening steps of \citetalias{Colella_1984}. We instead prevent new extrema in the reconstructed states by limiting the interface states at the faces of a given cell to the minimum and maximum of the cell-average values of cell under consideration and its two neighbouring cells along the axis of reconstruction, similar to the monotonicity constraint introduced by \cite{Mignone_2005}. This is followed by the extrema detection and overshoot correction step within each cell as described by \citetalias{Colella_1984}. In this step, the parabola assumed to exist across each cell is examined. If an `overshoot' (as defined by \citetalias{Colella_1984}) of the parabola is detected, we follow the original \citetalias{Colella_1984} prescription of performing linear reconstruction on the side of the cell affected by the overshoot. If an extremum is instead detected, rather than forcing the reconstruction to a constant value across the cell as done by \citetalias{Colella_1984}, we revert to performing a linear reconstruction within the affected cell, following \cite{Balsara_2017}. We note that these latter two steps only examine the interface values, and do not guarantee that the interface states lie within neighbouring cell-average values, and therefore the cell-average limiting carried out in the first step is not redundant. Any of these limiting steps may make the interface states discontinuous, with distinct states associated with each of the two cells adjacent to an interface.

We also implement reconstruction based on a piecewise-linear method (PLM) using the monotonized-central (MC) slope limiter \citep{VanLeer_1977}. We use PPM reconstruction by default, but allow PLM reconstruction via a compile-time option.

In some cases, especially in underresolved strong shocks, the previous steps do not provide sufficient dissipation to avoid oscillations. This problem was recognized by \citetalias{Colella_1984}, who proposed a shock flattening procedure in combination with a small amount of artificial viscosity. We find that this shock flattening procedure is not sufficient in multidimensional problems. Instead, we follow \cite{Miller_2002}, who generalize the \citetalias{Colella_1984} shock-flattening procedure for multidimensional hydrodynamics. Using this latter method, we find that no artificial viscosity is needed and we do not include any in our implementation.

Finally, in order to compute the flux of mass, momentum and energy between cells, we use the HLLC Riemann solver with the `primitive variable Riemann solver' wavespeeds and intermediate states \citep{Toro_2013}. We make the standard approximation that the face-average flux is the same as the face-centered flux, and therefore this step is also second-order accurate in spatial resolution. For each cell, the fluxes across each face are then added together to produce an unsplit spatial divergence term used by each stage of the Runge-Kutta integrator to advance the cell in time.

In multidimensional simulations, it has been long recognized that in strong grid-aligned shocks, the HLLC Riemann solver can unphysically amplify the so-called `carbuncle' instability \citep{Quirk_1994}. In astrophysical problems, this is most often encountered in strong explosions. Implementing additional dissipation in the form of artificial viscosity (e.g., \citealt{Gittings_2008}), the `H-correction' \citep{Sanders_1998}, or by adaptively switching to an HLL Riemann solver \citep{Harten_1983} for computing fluxes perpendicular to strong shocks (e.g., \citealt{Quirk_1994,Skinner_2019}) are possible solutions to this issue. In future work we plan to implement an adaptive procedure to fix the carbuncle instability via the latter method.

Future work may also include implementing an adaptive method to reduce the order of reconstruction in order to preserve density and pressure positivity in near-vacuum regions, such as the multidimensional optimal order detection (MOOD) method of \cite{Clain_2011}. An alternative solution may be to adaptively switch to an exact (iterative) Riemann solver depending on the flow conditions \citep{Toro_2013}.

\subsubsection{Radiation}
\label{sssec:radiation}

We solve the radiation transport subsystem (\autoref{eq:rad_energy}--\autoref{eq:rad_flux}, again omitting the terms on the right-hand side) in a similar method-of-lines fashion. Our approach is most similar to that of \cite{Skinner_2019}, who also evolve the radiation moment equations with a time-explicit method-of-lines approach; however, they do not use either PPM reconstruction or a reduced speed of light. Because even with the RSLA the signal speed for the radiation subsystem is substantially larger than for the hydrodynamic subsystem, we evolve the former explicitly in time with several radiation timesteps per hydrodynamic timestep. In the regime of applicability of the RSLA, this approach allows a much more computationally efficient solution to the radiation moment equations, due to the fact that explicit methods have a greater arithmetic intensity per byte of data, have simple memory access patterns and control flows (compared to implicit solvers), and do not require global communication across the computational domain in order to advance the solution in time. All these features are greatly beneficial on GPUs, where the ratio of floating-point arithmetic performance to memory bandwidth is typically greater than on CPUs.

We carry out each radiation subcycle using the same RK2-SSP integrator \citep{Shu_1988} that we use for hydrodynamics. We likewise use a finite volume representation of the radiation variables, with a PPM spatial reconstruction (or optionally, PLM) of the radiation energy density $E_r$ and reduced flux $\mathbf{f} = \mathbf{F}_r / cE_r$; the only difference in our procedures for hydrodynamics and radiation is that for radiation we do not employ a shock flattening procedure. There can exist unphysical radiation shocks when using local closures, since in general such closures make the radiation subsystem nonlinear, but there is no applicable shock flattening procedure to suppress this effect. We carry out reconstruction in terms of the reduced flux $\mathbf{f}$ rather than the absolute flux $\mathbf{F}_r$ in order to suppress unphysical fluxes $|\mathbf{F}_r| > cE$. This is effective in 1D problems, but in multidimensional problems, the magnitude of the radiation flux may still exceed $cE_r$, which is an unphysical state in which local closures cannot compute the Eddington factor at all. Reducing the order of reconstruction to first order (piecewise constant) when the interface states violate this constraint helps but does not eliminate the issue in all cases. For the purpose of computing the local closure only, we use rescale the flux such that $|\mathbf{F}_r| = cE_r$ whenever $|\mathbf{F}_r| > cE_r$.  For particularly difficult problems, especially in order to avoid unphysical instabilities in the propagation of non-grid-aligned optically-thin radiation fronts, we find that it is necessary to reconstruct the radiation variables using PLM reconstruction.

One drawback to upwind finite volume methods for radiation transport is that in naive form, they do not give the correct behavior for diffusive regions where the optical depth per cell is much greater than unity. This failure occurs because numerical diffusion dominates over physical diffusion when using upwind methods when the mean free path of photons is not resolved \citep{Lowrie_2001}. One common approach to fix this incorrect behavior is to modify the Riemann solver in the optically thick regime to reduce the upwind bias of the spatial derivative \citep{Audit_2002,Skinner_2019,Mezzacappa_2020}. However, this can lead to violations of causality (i.e., $|\mathbf{F}_r| > cE_r$) when the radiation flux is in the streaming regime \citep{Audit_2002}, which occurs especially at discontinuities in the opacity between optically-thin cells and optically-thick cells. The only apparent fix for this problem, which we adopt, is to disable the optical-depth correction in the Riemann solver for those cells where it produces a causality-violating state. We find that this condition is only activated when $f \rightarrow 1$, so it may not qualitatively affect the solution. However, we also advocate refining on the gradient in the optical depth per cell in order to resolve the boundary layers in such situations whenever it is computationally feasible.

For computing the flux of radiation quantities between cells, we use an HLL Riemann solver, with wavespeeds computed assuming the Eddington factors are fixed at the beginning of the timestep \citep{Balsara_1999}. This approach allows us to substitute different closure relations for the Eddington factors without requiring a modification of the Riemann solver, unlike previous implementations that are restricted to a single local closure (e.g., \citealt{Gonzalez_2007,Skinner_2013,Skinner_2019}). In principle, we could even use Eddington tensors computed via a short characteristics formal solution of the radiative transfer equation (e.g., \citealt{Davis_2012}), but we leave exploration of a non-local VET method to future work.

Our default closure relation for the Eddington tensor is the \cite{Levermore_1984} closure, which is derived by assuming that the radiation field is isotropic in some (unknown) reference frame and then computing a Lorentz transform from this reference frame to one in which the reduced flux $\mathbf{f}$ matches the value in the cell under consideration. This procedure leads to a radiation pressure tensor (e.g., \citealt{Gonzalez_2007,Rosdahl_2013,Skinner_2013})
\begin{align}
    \label{eq:M1_closure}
    \mathsf{P}_r = \left( \frac{1 - \chi}{2} \mathsf{I} + \frac{3\chi - 1}{2} \mathbf{\hat n} \mathbf{\hat n} \right) E_r
\end{align}
where $\mathsf{I}$ is the identity tensor, and the Eddington factor $\chi$ and the flux direction cosine $\mathbf{\hat n}$ are
\begin{align}
    \chi = \frac{3 + 4f^2}{5 + 2 \sqrt{4 - 3 f^2}} \, , \\
    \mathbf{\hat n} = \frac{\mathbf{F}_r}{|\mathbf{F}_r|} \, .
\end{align}
When the radiation flux is exactly zero, we drop the direction-dependent term in \autoref{eq:M1_closure}. By considering a coordinate system where the radiation flux is aligned with a coordinate axis, we see that $\chi$ is the component of the Eddington tensor in the direction of the radiation flux.

We emphasise that this is only one possible choice of closure, and a variety of alternative local closures exist \citep[e.g.,][]{Minerbo_1978,Levermore_1981}. We refer readers to \citet{Janka_1992} and \citet{Koerner_1992} for systematic comparisons to angle-dependent transport solutions for neutrinos, and \citet{Olson_2000} for comparisons to photon solutions. Because of its prominence in the neutrino transport literature, as well as marginally favorable performance on some test problems, we also provide an implementation of the \cite{Minerbo_1978} closure in addition to the default \citet{Levermore_1984} option. However, users can also implement any local closure of their choice simply by providing an implementation of a function that maps from the reduced flux $\mathbf{f}$ to the Eddington factor $\chi$ for their preferred closure. Doing so does not come at any cost in computational performance.

\subsubsection{Matter-radiation coupling}
\label{sssec:coupling}

Following the computation of the hyperbolic part of the radiation subsystem, we use an implicit method to evaluate the source terms (those appearing on the right-hand sides of \autoref{eq:hydro_continuity}--\autoref{eq:rad_flux}) for both the radiation and hydrodynamic subsystems; this update occurs once per radiation subcycle, and thus several times per hydrodynamic step. Since there are no spatial derivatives in these terms, each cell can be updated independently.

The radiation-matter coupling update occurs in three steps. The first is to handle the energy source terms $c\rho (\kappa_P a_r T^4 - \kappa_E E_r)$ that appear in \autoref{eq:hydro_energy} and \autoref{eq:rad_energy}. In the regime of problems to which we can apply the RSLA this term is often the stiffest, and we therefore update it using the backward-Euler implicit method of \cite{Howell_2003}, specialized to the case of a single material and extended to include a reduced speed of light. Let $E_g = E - \rho v^2/2$ be the gas internal energy, and let $E_g^{(t)}$ and $E_r^{(t)}$ be the gas internal energy and radiation energy at the end of the hyperbolic update, where the superscript $(t)$ indicates quantities evaluated at this point in the update cycle. We compute the new gas internal energy $E_g^{(t+1)}$ and radiation energy $E_r^{(t+1)}$, where $(t+1)$ indicates the state after accounting for the exchange term, by solving the implicit system
\begin{eqnarray}
    0 = F_G &\equiv &\left(E_g^{(t+1)} - E_g^{(t)}\right) + \left( \frac{c}{\hat c} \right) R^{(t+1)} \\
    0 = F_R &\equiv & \left(E_{r}^{(t+1)} - E_{r}^{(t)}\right) - \left( R + S \right)^{(t+1)},
\end{eqnarray}
where
\begin{equation}
    R \equiv \Delta t \rho \kappa_P (4 \pi B - \hat c E_r),
\end{equation}
$\Delta t$ is the radiation substep timestep, and $S$ is an optional source term that we include to allow, for example, addition of radiation by stellar sources. The quantities $F_G$ and $F_R$ are the residual errors in the gas energy and radiation energy, respectively.

To solve this system via Newton-Raphson iteration, we require the Jacobian matrix, the elements of which are
\begin{align}
    \frac{\partial F_G}{\partial E_g} & = 1 + \left( \frac{c}{\hat c} \right) \frac{\partial R}{\partial E_g} \, , \\
    \frac{\partial F_G}{\partial E_r} & = -c \Delta t \rho \kappa_P \, ,                                           \\
    \frac{\partial F_R}{\partial E_g} & = -\frac{\partial R}{\partial E_g} \, ,                                    \\
    \frac{\partial F_R}{\partial E_r} & = 1 + \hat c \Delta t \rho \kappa_P \, ,
\end{align}
where
\begin{align}
    \frac{\partial R}{\partial E_g} & = \frac{\rho \Delta t}{C_v} \left[ \kappa_P \frac{\partial B}{\partial T} + \frac{\partial \kappa_P}{\partial T} \left( 4\pi B - \hat c E_r^{t+1} \right) \right] \, ,
\end{align}
and $C_v$ is the gas total heat capacity at constant volume. From the Jacobian, we can write the change in radiation and gas temperature for each iterative update as
\begin{align}
    \label{eq:delta_er}
    \Delta E_r & = -\frac{F_R + \eta F_G}{ \frac{\partial F_R}{\partial E_r} + \eta \frac{\partial F_G}{\partial E_r} } \, , \\
    \label{eq:delta_eg}
    \Delta E_g & = -\frac{F_G + \Delta E_r \frac{\partial F_G}{\partial E_r}}{ \frac{\partial F_G}{\partial E_g} } \, .
\end{align}
where $\eta \equiv - (\partial F_R/\partial E_g) ( \partial F_G/\partial E_g )^{-1}$. We repeatedly apply \autoref{eq:delta_er} and \autoref{eq:delta_eg} to the radiation and gas energies until the system converges. \citet{Howell_2003} leave unspecified the convergence criteria they use for their solver. After experimenting with several possibilities, we decide to stop the Newton-Raphson iterations when the residuals $F_R$ and $F_G$ satisfy
\begin{align}
    \left| \frac{F_G}{E_{\text{tot}}} \right|                  & < \epsilon \, \, \text{and} \\
    \left| \frac{c}{\hat c} \frac{F_R}{E_{\text{tot}}} \right| & < \epsilon \, ,
\end{align}
where
\begin{align}
    E_{\text{tot}} & \equiv E_g^{(t)} + \frac{c}{\hat c} \left( E_r^{(t)} + S \right) \, .
\end{align}
When $\hat c = c$, $E_{\text{tot}}$ is the total (internal gas plus radiation) energy at the end of the timestep. By default, the relative tolerance $\epsilon$ is set to $10^{-10}$. We find that larger tolerances produce unacceptably inaccurate solutions for many problems. In especially stiff problems, it may be necessary to reduce the tolerance to the order of machine precision for double-precision floating point arithmetic ($\sim 10^{-15}$). If the solver exceeds a specified maximum number of iterations (400 by default) without converging, the code prints an error message and exits. Convergence failure usually occurs only when the intial timestep has not been sufficently reduced compared to the CFL timestep at the start of a simulation.

Once the Newton-Raphson iterations have converged, we have obtained the updated gas internal energy and radiation energy, and we proceed to the next two steps of updating the coupling terms. We first update the radiation and gas momenta, accounting for the coupling term $\rho \kappa_F \mathbf{F}_r/c$. To do so, we compute the flux mean opacity $\kappa_F$ using the updated gas temperature. Then, following \cite{Skinner_2019}, we use a backward-Euler discretization of the radiation flux source term (modified to include a reduced speed of light):
\begin{align}
    \vc{F}_r^{(t+1)} = \frac{\vc{F}_r^{(t)}}{1 + \rho \kappa_F \hat c \Delta t} \, .
\end{align}
In order to ensure momentum conservation when $\hat c = c$, we apply the difference in radiation flux in an equal and opposite manner to the gas momenta (as advocated by \citealt{Skinner_2019}):
\begin{align}
    \Delta \vc{F}_r         & \equiv \vc{F}_r^{(t+1)} - \vc{F}_r^{(t)} \, ,                   \\
    {(\rho \vc{v})}^{(t+1)} & = {(\rho \vc{v})}^{(t)} - \frac{\Delta \vc{F}_r}{\hat c c} \, .
\end{align}

The final step is to compute the work done by the radiation force on the gas. Since we are evolving the conserved variables, this term cannot be computed explicitly as written in \autoref{eq:rad_energy} without causing a significant error in the gas internal energy when the radiation force is stiff. We instead compute this term as the difference in gas kinetic energy over the timestep $\Delta E_{\text{kin}}$, then add this quantity to the total gas energy and subtract this quantity from the radiation energy:
\begin{align}
    E^{(t+1)}   & \leftarrow E_g^{(t+1)} + \left( E_{\text{kin}}^{(t)} + \Delta E_{\text{kin}} \right) \, , \\
    E_r^{(t+1)} & \leftarrow E_r^{(t+1)} - \left( \frac{\hat c}{c} \right) \Delta E_{\text{kin}} \, ,
\end{align}
where $E^{(t+1)}$ denotes the total gas energy at the end of the timestep. This completes the update for all radiation-matter coupling terms.

\subsubsection{Level synchronization procedure}
\label{sssec:sync}

As explained by \cite{Berger:1989}, in an AMR calculation it is necessary to adjust the solution on the coarse AMR level following the solution on any refined level in order to maintain conservation of the evolved quantities (e.g., mass, momentum, energy). For hyperbolic equations evolved explicitly in time, this is traditionally done by saving the flux at the coarse-fine grid boundary in a `flux register' for both the flux computed on the fine level and the flux computed on the coarse level. In general, these fluxes are different due to the differing stencil used on the coarse and fine levels, and without correction, this would lead to a loss of conservation of energy (and any other conserved quantities). The flux register stores this mismatch, and in the synchronization step, adds the missing mass, momentum, or energy to the cells on the coarse level immediately adjacent to the coarse-fine boundary.

As noted by \cite{Howell_2003}, an implicit radiation update has additional difficulties in ensuring conservation, since radiation can propagate much further than a single grid cell on the coarse grid. Our radiation update is fully explicit, but we would like to advance each AMR level on the hydrodynamic timescale, rather than on the radiation timescale, so we have a similar long-range signal propagation difficulty. \cite{Rosdahl_2013} outline three possible solutions to the problem: i) perform the radiation solve after each coarse hydrodynamic step, keeping subcycling-in-time on refined levels (which would be very inaccurate), ii) use a single global timestep for all AMR levels, which allows one to advance the radiation solution on all levels in each radiation substep (which would be very computationally expensive, since in our applications of interest, the global timestep is typically limited by the timestep of the highest-resolution level) or iii) restrict the timestep for each level to the minimum of the radiation and hydrodynamic timesteps. In our code, we set the coarse timestep such that the number of radiation substeps per level is limited to a maximum value $N_{\text{sub,max}}$ in order to minimize the signal propagation distance from the coarse-fine boundaries. The flux mismatch at the coarse-fine boundaries is added to the immediately adjacent cells on the coarse grid at the end of each level advance. When $N_{\text{sub,max}} = 1$, our solution is identical to the flux synchronization method used in the \textsc{Ramses} AMR code \citep{Rosdahl_2013}. However, as a default we set the parameter $N_{\text{sub,max}}$ to $10$, which appears to be sufficient to avoid significant discontinuities in the radiation energy and flux at coarse-fine boundaries, but still allows for significant subcycling and thus a substantially lower computational cost. We use this value for all test problems shown in this work, but users are able to set this parameter as desired for either greater efficiency or greater consistency at refinement boundaries. When this parameter is too large, however, it is possible for the coarse level to fail to maintain positivity of the radiation energy or causality of the radiation flux.

\section{Test problems}
\label{section:tests}

We now proceed to describe a series of tests that we have conducted to verify \quokka's accuracy and convergence characteristics, starting with tests of the hydrodynamic subsystem (\autoref{ssec:hydro_tests}), followed by tests of the radiation transport and radiation-matter exchange subsystems (\autoref{ssec:radiation_tests}), and concluding with tests of coupled radiation hydrodynamics (\autoref{ssec:radhydro_tests}).

Additional example problems and an automated test suite of $20$ test problems with checks against exact solutions are included with \textsc{Quokka}'s source code. We run this test suite for each commit and pull request in our GitHub repository. While continuous integration tests such as ours cannot guarantee bug-free software, this practice has flagged and prevented the introduction of several bugs during the development of \textsc{Quokka}. In order to maintain high software quality, we also run the commercial static code analyzer \textsc{SonarQube}\footnote{Available from SonarSource S.A, Switzerland via \url{https://www.sonarqube.org}. We have detected several bugs affecting solution correctness in other hydrodynamics codes using this tool.} on every commit and pull request.

\subsection{Hydrodynamics}
\label{ssec:hydro_tests}

For all our hydrodynamics tests, we disable the radiation portion of the code. These tests evaluate the hydrodynamic transport solver in isolation.

\subsubsection{Sound wave}
We compute the propagation of a sound wave in one dimension in order to measure the convergence of our numerical method to the exact solution as a function of spatial resolution, following the test described by \cite{Stone_2008}. With $\vc{U}$ denoting the vector of conserved variables $(\rho, \rho v_x, \rho v_y, \rho v_z, E)$, we initialize the simulations with the initial state $\vc{U}_0 + \delta \vc{U}$, where
\begin{align}
    \delta \vc{U} = A \vc{R} \sin(2\pi x)
\end{align}
where $\vc{R} = (1,-1,1,1,1.5)$ is the right eigenvector of the linearized hydrodynamic system, and $\vc{U}_0$ is the background state with density $\rho = 1$, velocity $\vc{v} = 0$, and pressure $P = 1/\gamma$. We set the adiabatic index $\gamma = 5/3$ and the wave amplitude $A = 10^{-6}$. We simulate a periodic domain $x= 0$ to 1 and evolve the system for one wave period, allowing us to compute the error of the solution by direct comparison of the initial conditions and the final state of the simulation. We define the error vector
\begin{align}
    (\Delta \vc{U})_{k} = \frac{1}{N_x} \sum_{i} \left| U_{i,k} - U_{i,k}^{0} \right|
    \label{eq:convergence_error}
\end{align}
where $k$ denotes a component of each state vector, $\vc{U_i}$ is the vector of conserved variables in cell $i$ at the final timestep, and $\vc{U_i}^{0}$ is the vector of conserved variables in cell $i$ in the initial conditions. Each component $|\Delta U|_k$ is therefore the $L_1$ norm of the error of a component of the solution state.  We assess the accuracy of the solution based on the root-mean-square (rms) of the components of this error vector, denoted $||\Delta \vc{U}||$.

\begin{figure}
    \includegraphics[width=\columnwidth]{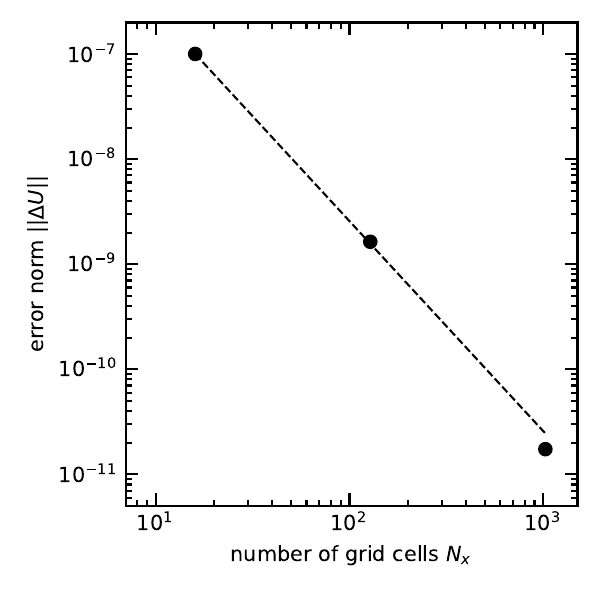}
    \caption{The error $||\Delta \vc{U}||$ in the solution (\autoref{eq:convergence_error}) for a sound wave as a function of spatial resolution per wavelength $N_x$. Black circles show numerical results, and the dashed line is a power law that scales as $N_x^{-2}$ normalised to the observed error at the smallest $N_x$.}
    \label{fig:sound_wave}
\end{figure}

We run simulations using PPM reconstruction and a CFL number of $0.1$, using grid sizes from $N_x = 16$ to $N_x = 1024$. We show the error norm as a function of resolution in \autoref{fig:sound_wave}. For $N_x = 16$, we obtain $||\Delta \vc{U}|| = 1.0 \times 10^{-7}$, for $N_x = 128$, we obtain $||\Delta \vc{U}|| = 1.6 \times 10^{-9}$, and for $N_x = 1024$, we obtain $||\Delta \vc{U}|| = 1.7 \times 10^{-11}$. The results for our code are in excellent agreement with those from the \textsc{Athena} hydrodynamic solver (Figure 7 of \citealt{Stone_2008}). The $N_x^{-2}$ scaling of the error norm indicates that our hydrodynamic solver converges at second order in spatial resolution, as expected from the formal order of accuracy of the method.

\subsubsection{Contact discontinuity}
The HLLC Riemann solver has the property that it can resolve an isolated stationary contact discontinuity with infinite resolution \citep{Toro_2013}. The HLL solver, on the other hand, introduces a large amount of numerical diffusion for this problem (see Figure 10.20 of \citealt{Toro_2013}). To verify that our hydrodynamic implementation can maintain a perfect contact discontinuity, we simulate a system where the initial conditions have a left and right state separated with a discontinuity at $x = 0.5$. The left state is $\rho_L = 1.4$, $p_L = 1.0$, and the right state is $\rho_R = 1.0$ and $p_R = 1.0$. Since this is a pure contact discontinuity, the solution should not evolve from the initial state. We set the velocity to zero, and use an adiabatic index $\gamma = 1.4$. We evolve the solution numerically until $t = 2$. The error with respect to the correct solution is exactly zero.

\subsubsection{Stationary shock tube}
\label{section:shocktube}
Our next test is a stationary shock tube, which we set up using the parameters suggested on the website of F.X. Timmes\footnote{\url{http://cococubed.asu.edu/code_pages/exact_riemann.shtml}}. This shock tube problem is substantially more difficult to solve than the standard \cite{Sod_1978} shock tube test due to the larger jump in pressure and density at the discontinuity. We initialize left and right states with a discontinuity at $x_0 = 2$, with the left state $\rho_L = 10$, and $p_L = 100$, and the right state $\rho_R = 1$ and $p_R = 1$. The initial velocity is zero. We evolve the solution using a CFL number of 0.1 until $t = 0.4$ on a grid of 1000 cells on the domain $[0, 5]$. We use a small CFL number since the wave structure at the discontinuity creates waves that propagate faster than the linearized Roe eigenvalues would predict.

\begin{figure}
    \includegraphics[width=\columnwidth]{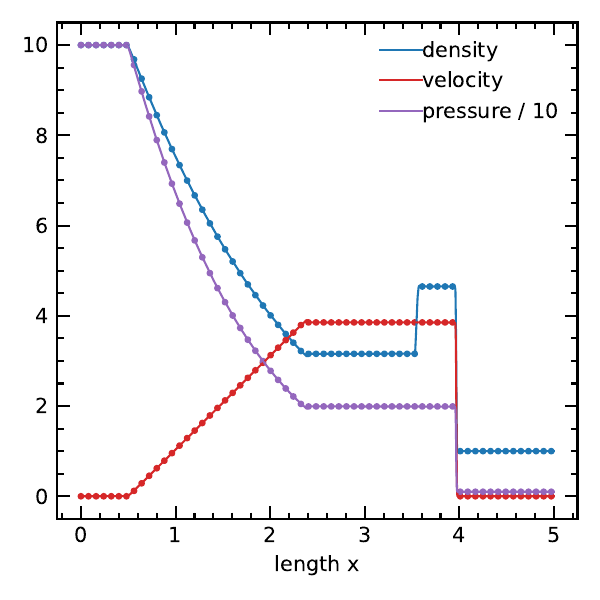}
    \caption{Density, velocity, and pressure profiles for the stationary shock tube test (\autoref{section:shocktube}). Lines show the \quokka~simulation result. For comparison, we show the exact solutions for density, velocity, and pressure as circles of the same color.}
    \label{fig:shocktube}
\end{figure}

We show \quokka's results for this test in \autoref{fig:shocktube}. As for the sound wave test, we compute the $L_1$ error norm for each of the conserved variables, and then compute the root-mean-square of those error norms. The rms $L_1$ error norm divided by the rms norm of the exact solution is $1.12 \times 10^{-3}$. Inspecting the solution in \autoref{fig:shocktube}, we see that the agreement between the exact solution and the numerical solution is very good. The only noticeable differences are small oscillations near discontinuities in the derivative of the solution at $x \approx 2.4$ and near the density discontinuity at $x \approx 3.6$. We find that the shock flattening method of \cite{Miller_2002} is essential to produce a reasonable solution to this problem. Without it, we find unacceptably large post-shock oscillations (not shown).

\subsubsection{`LeBlanc' test}
\label{section:leblanc}
\begin{figure}
    \includegraphics[width=\columnwidth]{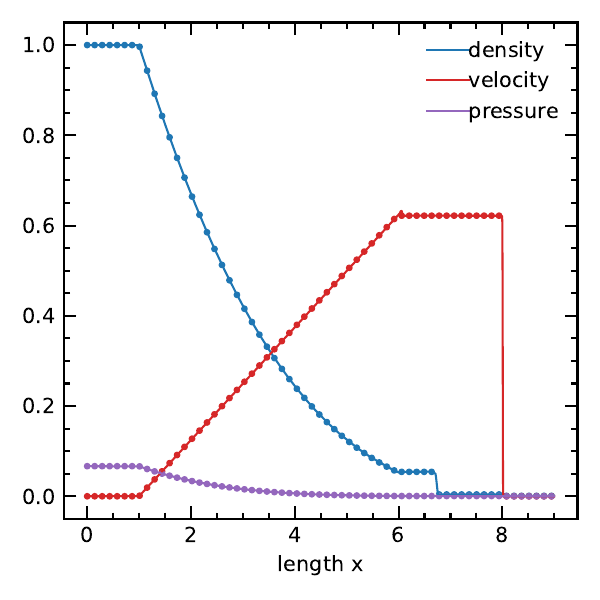}
    \includegraphics[width=\columnwidth]{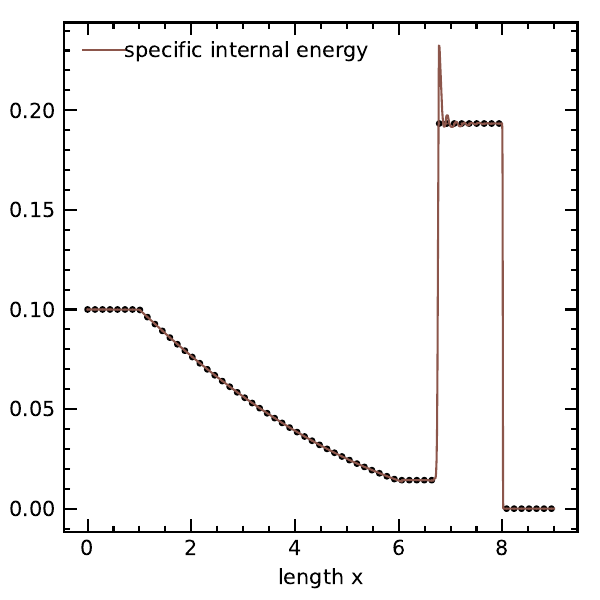}
    \caption{\emph{Upper:} Density, velocity, and pressure profiles for the `LeBlanc' shock tube test problem (\autoref{section:leblanc}). \emph{Lower:} The specific internal energy. In both panels, solid lines show the \quokka~simulation result, and circles in corresponding colours show the exact result.}
    \label{fig:leblanc}
\end{figure}

We next carry out the `LeBlanc' shock tube test, originally published by \cite{Benson_1992} and further described by \cite{Pember_2001}. In this problem, we initialize a left state with $\rho_L = 1$ and $p_L = \frac{2}{3} \times 10^{-1}$, and a right state with $\rho_R = 10^{-3}$ and $p_R = \frac{2}{3} \times 10^{-10}$. We set the initial velocity to zero and use an adiabatic index $\gamma = 5/3$. This is an extreme shock tube that far exceeds any shock that may be encountered in any conceivable application, featuring a pressure jump of nine orders of magnitude, and is therefore an excellent test problem.  We evolve this simulation until $t = 6$ using a grid of 2000 cells and a CFL number of 0.1. The resulting state is shown in \autoref{fig:leblanc}. \cite{Pember_2001} highlight the difficulty of obtaining the correct specific internal energy in the solution for this test, but we find that \quokka~produces the correct shock location and specific internal energy, with the exception of a small overshoot at the shock location. The use of shock flattening is essential for this problem. Overall, the performance of our code on this problem is excellent.

\subsubsection{Wave-shock interaction (Shu-Osher) test}
\label{section:shu-osher}
\begin{figure}
    \includegraphics[width=\columnwidth]{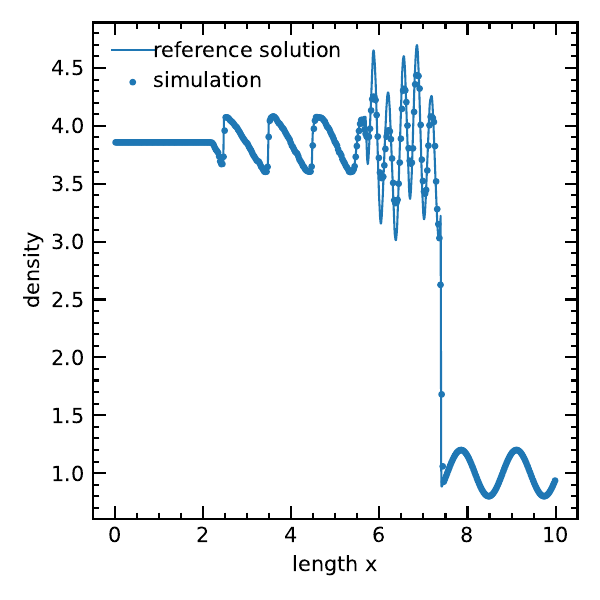}
    \caption{Density as a function of position for the Shu-Osher wave-shock interaction problem (\autoref{section:shu-osher}). Here we show a comparison between the \quokka~solution (shown as circles) and a reference solution (shown as a solid line) computed using \textsc{Athena++} \citep{Stone_2020}.}
    \label{fig:shu_osher}
\end{figure}

We show the Shu-Osher test in \autoref{fig:shu_osher}. Following the description of \cite{Shu_1989}, the initial conditions are, on the left side,
\begin{align}
    \rho_L(x) & = 3.857143 \, , \\
    v_L(x)    & = 2.629369 \, , \\
    P_L(x)    & = 10.33333 \, ,
\end{align}
and the right-hand state is
\begin{align}
    \rho_R(x) & = 1 + 0.2 \sin(5x) \, , \\
    v_R(x)    & = 0 \, ,                \\
    P_R(x)    & = 1 \, .
\end{align}
We compute a reference solution using \textsc{Athena++} \citep{Stone_2020} with the VL2 integrator, PPM reconstruction in the characteristic variables, and the HLLC Riemann solver on a grid of 1600 cells. Our solution is computed using PPM reconstruction (in the primitive variables), the RK2-SSP integrator, and the HLLC Riemann solver on a grid of 400 cells. The agreement is very good, with comparable resolution of the high-frequency features to the third-order essentially-non-oscillatory (ENO) scheme of \cite{Shu_1989}. When PLM reconstruction is used instead for the same number of grid cells, the high-frequency features are aliased (not shown; see also Figure 14 of \citealt{Shu_1989}), indicating a significantly higher effective resolution for PPM-based methods even in the presence of shocks.

\subsubsection{Slow-moving shock}
\label{section:slowshock}
\begin{figure}
    \includegraphics[width=\columnwidth]{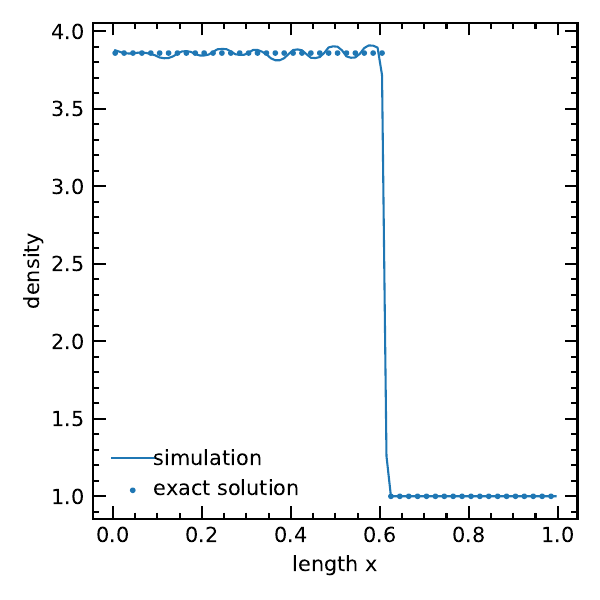}
    \caption{Density profiles for the slow-moving shock test problem (\autoref{section:slowshock}). We show a comparison between the \quokka~(solid line) and exact solutions (circles).}
    \label{fig:sms}
\end{figure}

We show a slow-moving shock in \autoref{fig:sms} using the parameters from \cite{Jin_1996}, where $\rho_L = 3.86$, $(\rho v)_L = -3.1266$, and $E_L = 27.0913$, and the right-side state $\rho_R = 1.0$, $(\rho v)_R = -3.44$, and $E_R = 8.4168$, with $\gamma = 1.4$. This corresponds to the shock jump moving to the right with a velocity $v_{\text{shock}} = 0.1096$. For a CFL number of $0.2$, this corresponds to the shock taking $\sim 250$ timesteps to move across a single cell. This may not be a common scenario for our applications, but it may occur in a protostellar accretion shock, for instance. The quality of the solution is again significantly improved by shock flattening. The post-shock oscillations for slow-moving shocks may still be present with first-order reconstruction \citep{Jin_1996,Lee_2011}, so it is difficult to completely eliminate. We also find that adding a small amount of artificial viscosity does not significantly reduce the oscillations. A modification to PPM reconstruction based on a characteristic wave decomposition succeeds in significantly reducing this oscillation \citep{Lee_2011}, which we may consider implementing in a future version of the code.

\subsubsection{Strong rarefaction}
\label{section:rarefaction}
\begin{figure}
    \includegraphics[width=\columnwidth]{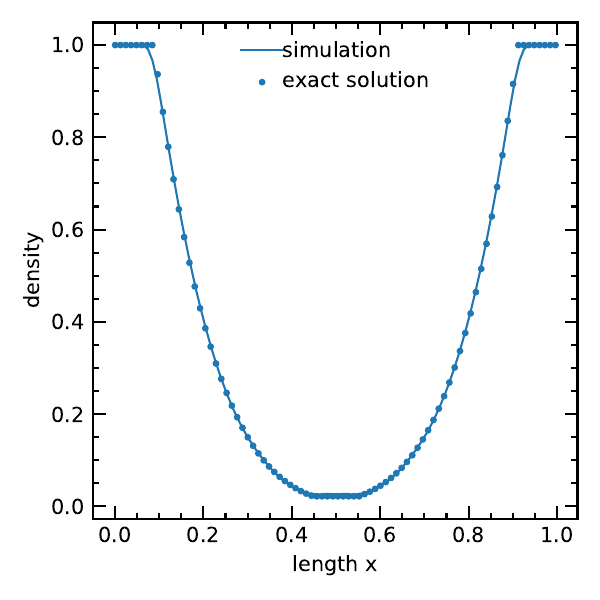}
    \includegraphics[width=\columnwidth]{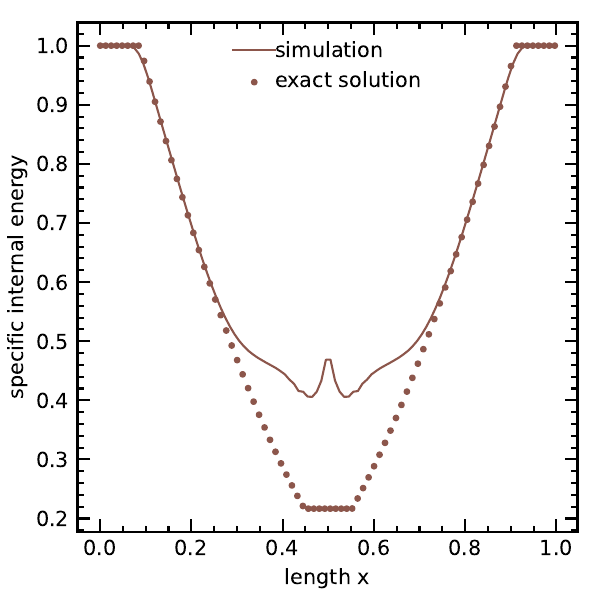}
    \caption{Profiles of density (top) and specific internal energy (bottom) for the strong rarefaction test problem (\autoref{section:rarefaction}). In both panels we show a comparison between the \quokka~solution computed using the default HLLC Riemann solver (solid line) and a solution computed using an exact, iterative Riemann solver (circles).}
    \label{fig:vacuum}
\end{figure}

We next test the performance of our code on the 1-2-3 problem of \cite{Einfeldt_1991}, which features a strong rarefaction and is designed to induce failures in approximate Riemann solvers. The initial conditions consist of left and right states with equal density and energy, $\rho_L = \rho_R = 1$ and $E_L = E_R = 3$, and equal magnitude but oppositely-directed velocities, $(\rho v)_L = -2$, $(\rho v)_R = 2$. We evolve the system to $t=0.15$, using a CFL number $0.8$ and a grid of $N_x = 100$ cells, and show the resulting state in \autoref{fig:vacuum}. We obtain the exact solution to which we compare the \quokka~result using an exact Riemann solver. We find that the solution for the density profile is very close to the exact solution, except for a small discrepancy at the lowest density near $x = 0.5$. However, the most difficult aspect of this problem is obtaining the correct specific internal energy. Our results compare favorably with other solutions obtained with approximate Riemann solvers, where factors of two or three errors are obtained near $x \approx 0.5$ \citep{Toro_2013}. Obtaining the correct specific internal energy in the lowest density part of the flow may require the adaptive use of an exact Riemann solver in near-vacuum regions. Nonetheless, our code is stable and well-behaved for this problem.

\subsubsection{Kelvin-Helmholtz instability}
\begin{figure}
    \begin{center}
        \includegraphics[width=0.85\columnwidth]{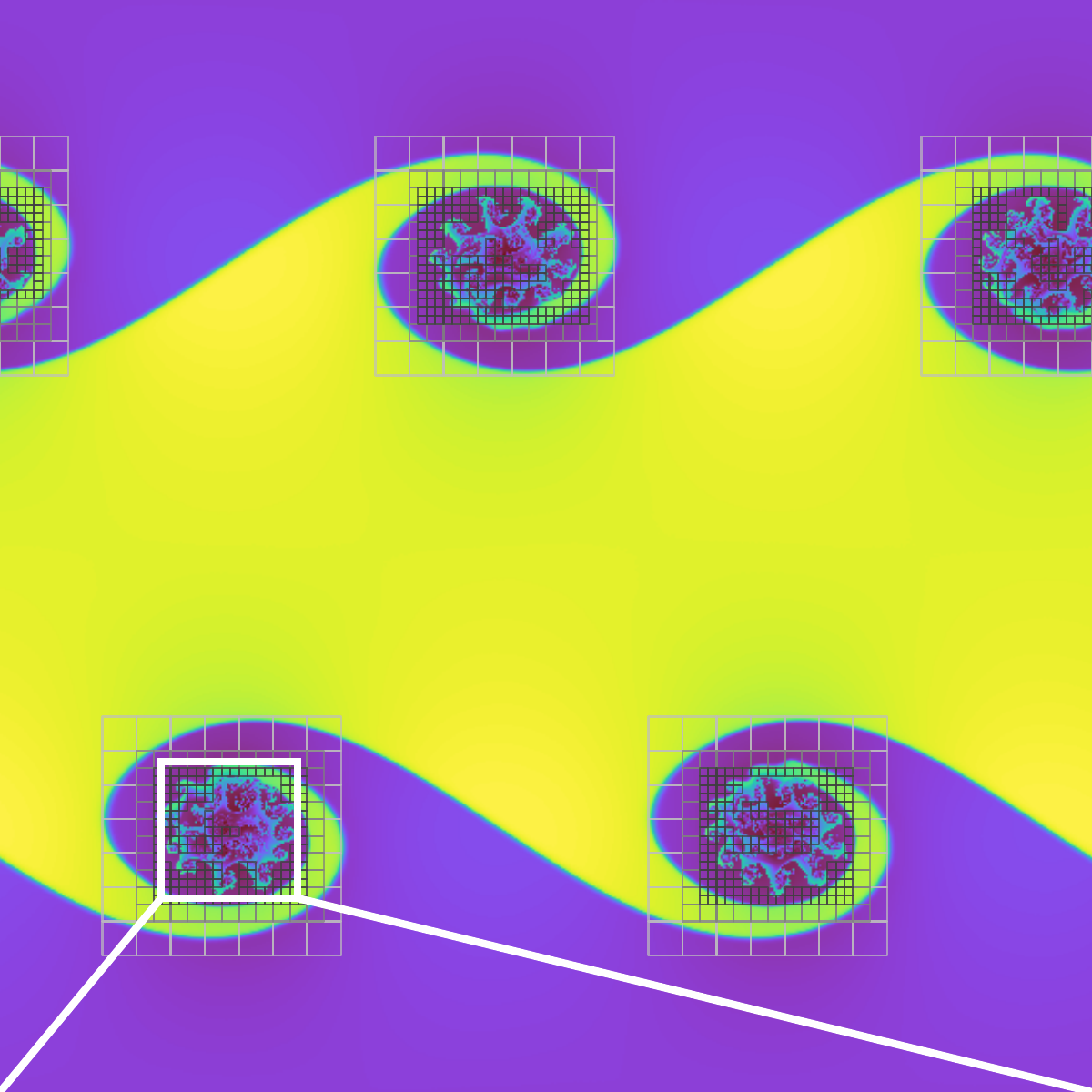}
        \includegraphics[width=0.85\columnwidth]{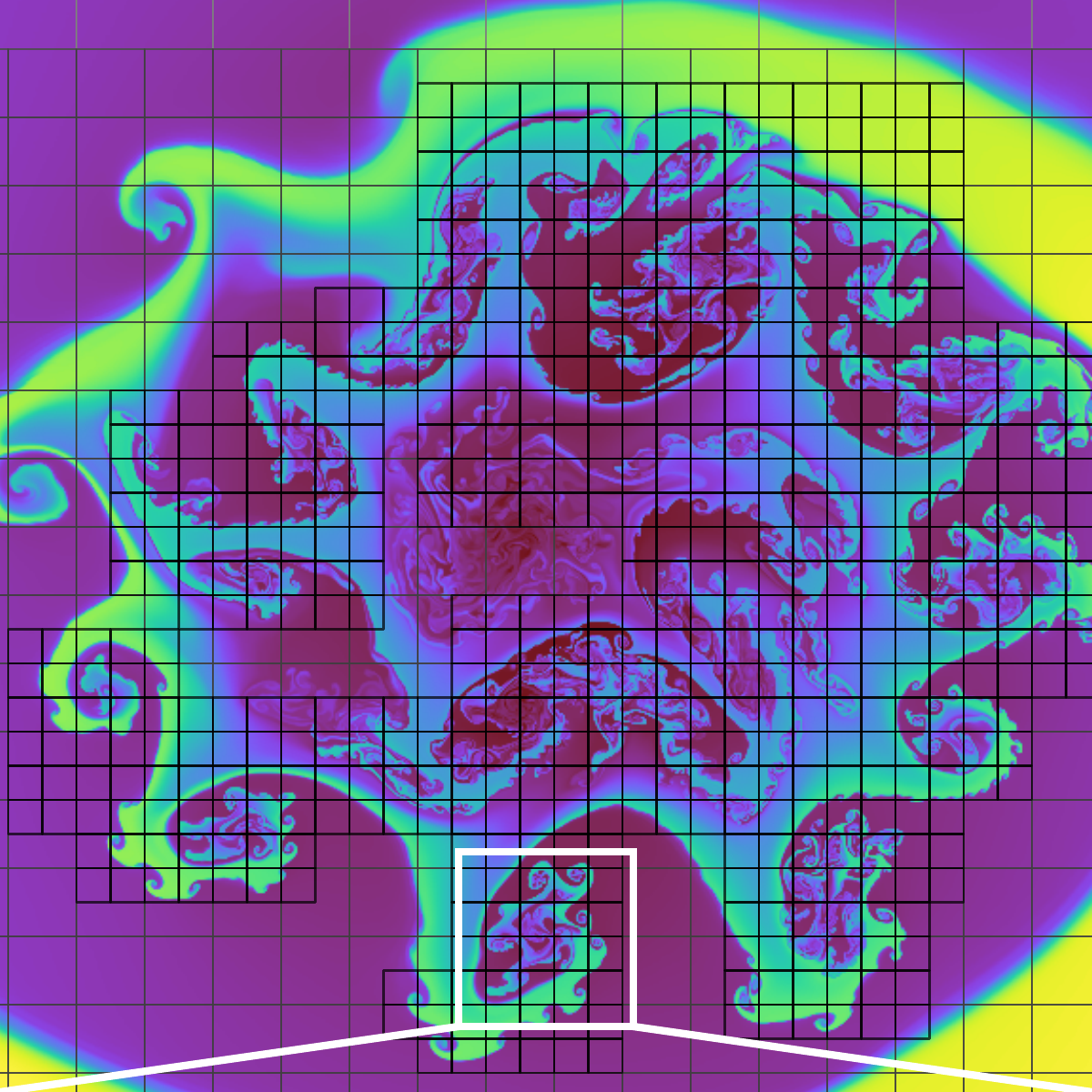}
        \includegraphics[width=0.85\columnwidth]{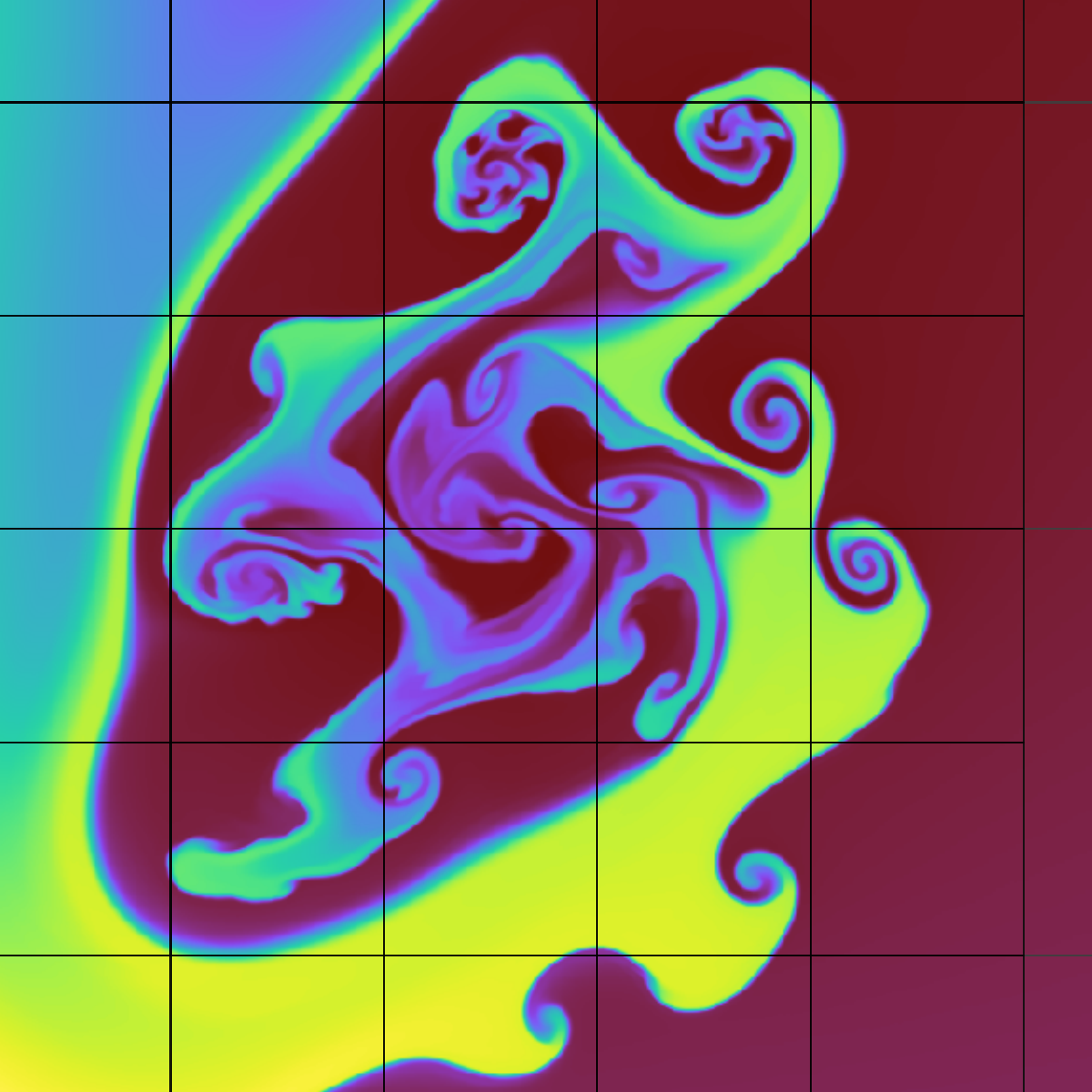}
    \end{center}
    \caption{A simulation of the Kelvin-Helmholz instability with 4 levels of refinement. The top panel shows the full simulation domain, and the lower two panels show-successive zoom-ins on parts of the domain. Grid boundaries are shown for levels $l \geq 2$. Color shows density.}
    \label{fig:kh_zoom}
\end{figure}

To show the performance of \textsc{Quokka} in two dimensions, we simulate the Kelvin-Helmholtz instability created by counter-propagating gas flows with a shear layer between them. The purpose of this test is to illustrate the ability of the code to maintain the contact discontinuity between the flows as Kelvin-Helmholz rolls develop, even as we add deep AMR nesting. Following \cite{Stone_2020}, we use a two-dimensional periodic box on the domain $[0, 1]$ along each axis with density and velocity given by:
\begin{align}
    \rho & =  1.5 - 0.5 \tanh(\tilde y / L) \, ,                           \\
    v_x  & = 0.5 \tanh(\tilde y / L) \, ,                                  \\
    v_y  & = A \cos(4\pi (x - x_0)) \, \exp(-{\tilde y}^2 / \sigma^2) \, ,
\end{align}
where $x_0 = 0.5$, $y_0 = 0.5$, $\tilde y = |y - y_0| - 0.25$, the shearing layer thickness $L = 0.01$, $\sigma = 0.2$, and perturbation amplitude $A = 0.01$. The initial pressure is uniform with $P = 2.5$ and we adopt an adiabatic index $\gamma = 1.4$. We enable AMR, with cells tagged for refinement if the relative density gradient on either side of the cell in either direction exceeds $0.2$, and we allow up to four levels of refinement on top of a base grid size of $2048^2$. Thus the peak resolution of the calculation is $32,768^2$. Each local AMR grid has a uniform size of $128^2$. We evolve the system to $t = 1.5$ with a CFL number of $0.4$, and show the resulting numerical solution in \autoref{fig:kh_zoom}. We are able to carry out this calculation on a single GPU in $\sim 4.5$ hours of wallclock time. While there appears to be no converged solution to this problem without explicit dissipation, we find that our hydrodynamic solver is able to resolve the Kelvin-Helmholz rolls with very little dissipation and with significant small-scale structure caused by secondary instabilities, as expected for inviscid simulations \citep{Lecoanet_2016}. There are no visible artifacts at resolution boundaries.

\subsubsection{Liska-Wendroff Implosion}
\begin{figure}
    \includegraphics[width=\columnwidth]{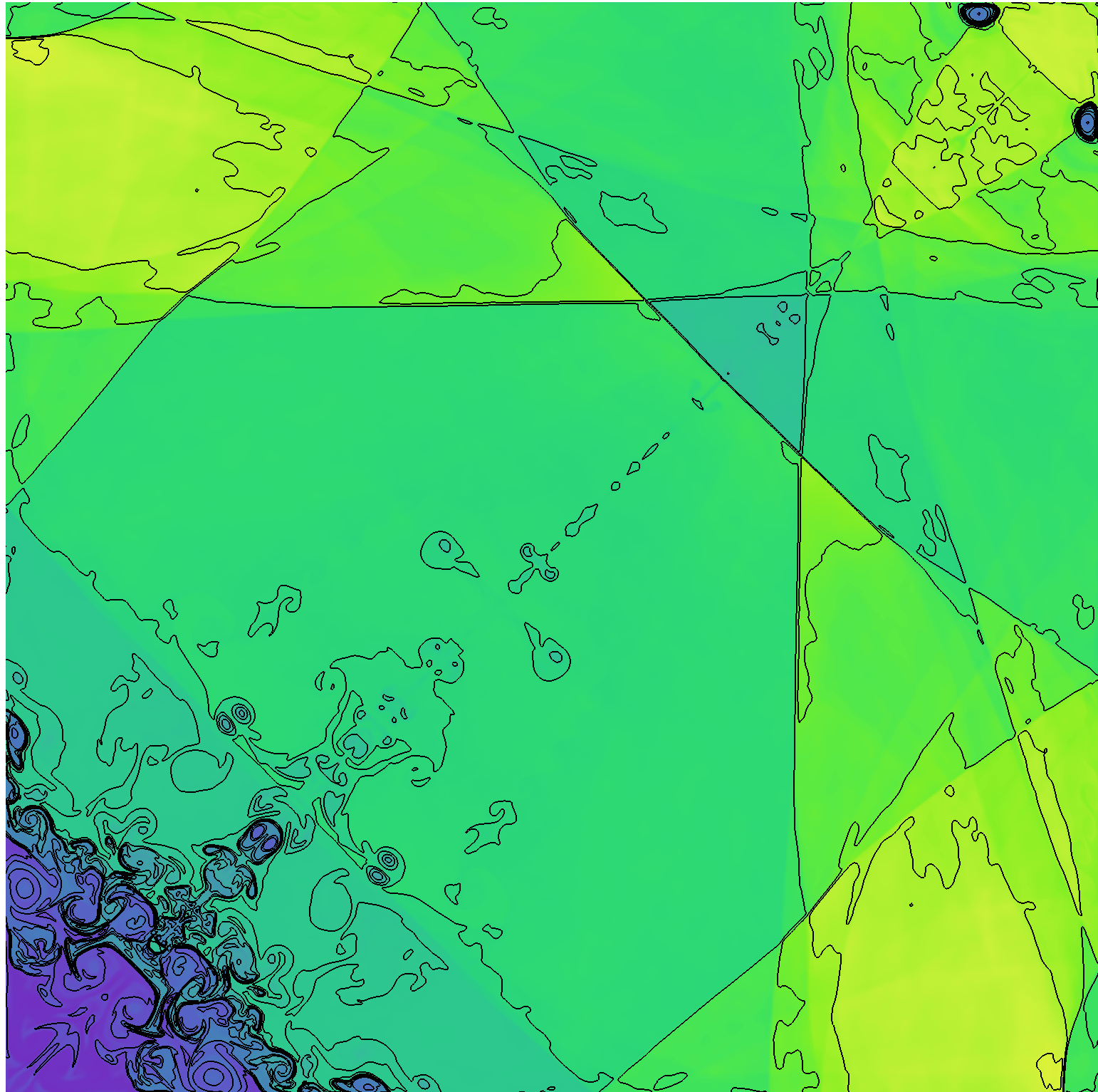}
    \caption{The two-dimensional implosion test \citep{Liska_2003} on a $1024^2$ grid at $t = 2.5$. The density is shown with 16 equally-spaced contours between 0.4 and 1.1, with the colormap showing the density over the same range. A  thin jet shoots along the $x$-$y$ diagonal. The solution is exactly symmetric.}
    \label{fig:implosion}
\end{figure}

We next present our results for the so-called Liska-Wendroff implosion test \citep{Hui_1999,Liska_2003}. This problem consists of the square domain $[0, 0.3]^2$, with an inner region $x+y \leq 0.15$ and an exterior region where $x + y > 0.15$ for an ideal gas with adiabatic index $\gamma = 1.4$. The inner region has initial density $\rho = 0.125$ and pressure $P = 0.14$ and the outer region begins with density $\rho = 1$ and pressure $P = 1$. We simulate the subsequent evolution to $t=2.5$ on a uniform grid of $1024^2$ cells with reflecting boundary conditions with a CFL number of $0.4$. These initial conditions lead to a shock directed toward the origin, which is then reflected many times by the upper and right walls before finally converging in a jet traveling away from the origin along the diagonal $x=y$, as shown in \autoref{fig:implosion}. \cite{Liska_2003} note that only codes that discretely preserve symmetry between x- and y-directions successfully produce the jet. In order to recover the jet in \quokka, we found it necessary to code the RK2-SSP integrator so that the fluxes in the x- and y-direction are added in an exactly symmetrical manner for each stage of the update. Additionally, when running the problem on NVIDIA GPUs, we preserve this symmetry only if we disable fused multiply-add (FMA) operations via the \texttt{nvcc} compiler option \texttt{fmad=false}, since the compiler otherwise breaks the symmetry expressed in the source code between the x- and y-direction fluxes. With this compiler option, \quokka~exactly preserves symmetry along the diagonal and successfully recovers the jet.

\subsection{Radiation}
\label{ssec:radiation_tests}

For our radiation tests we disable the hydrodynamic part of the code and only use the radiation transport and gas-radiation exchange updates. These tests evaluate the accuracy of those portions of the code.

\subsubsection{Marshak wave}
\label{ssec:marshak}
We next compute a Marshak wave \citep{Marshak_1958}. The problem consists of a uniform gas with a constant density $\rho = 10 \gm \cm^{-3}$ and constant opacities $\kappa_P = \kappa_R = 577 \cm^{2} \gm^{-1}$. The gas has a uniform initial temperature of $10^4 \Kelvin$, but at $t=0$ we impose on the left-hand side of the domain a boundary condition consisting of a half-isotropic flux with a radiation temperature of $3.481334 \times 10^6 \Kelvin$. The radiation drives a wave of heat into the gas. Following \cite{Su_1996}, we set the gas heat capacity at constant volume $C_v$ so a functional form that makes it possible to linearize the matter-radiation coupling terms, and thus obtain a semi-analytic solution:
\begin{align}
    C_v & \equiv \frac{\partial E_{\text{int}}}{\partial T} = \alpha T^3 \, ,
    \label{eq:heat_capacity}
\end{align}
where $E_{\text{int}} = (\alpha / 4) \, T^4$, $\alpha = 4 a_r / \epsilon$ and $\epsilon = 1$. With this heat capacity, \cite{Su_1996} obtain a semi-analytic quadrature solution of the radiation diffusion equation for this problem as a function of $\epsilon$. We evolve the solution until time $t = \tau / (\epsilon c \rho \kappa)$ where $\tau = 10$, using a simulation domain on the interval $[0 \cm, 3.466205 \times 10^{-3}\cm]$ resolved by grid of $N_x = 400$ cells. We note that this implies an optical depth per cell of $\tau_{\text{cell}} \approx 0.05$, so this problem does not test the accuracy of our code in the asymptotic diffusion limit (where $\tau_{\text{cell}} \gg 1$; the accuracy in this limit is instead tested via the radiation pressure tube problem in \autoref{section:radtube}). We do not use a reduced speed of light for this test.

\begin{figure}
    \includegraphics[width=\columnwidth]{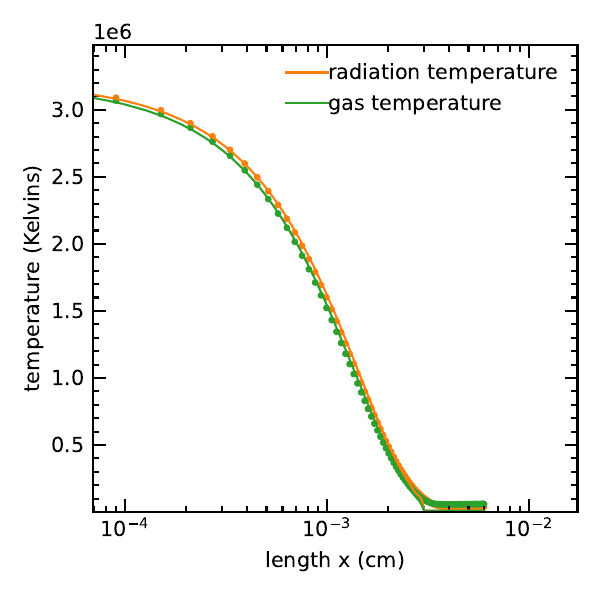}
    \caption{A Marshak wave test problem (\autoref{ssec:marshak}). The radiation temperature and gas temperature computed by \quokka~are shown as solid lines, while the analytic solution for the diffusion approximation is shown as circles.}
    \label{fig:marshak}
\end{figure}

Since we solve the moment equations, rather than just the diffusion equation, we do not expect our numerical solution to agree with the \citeauthor{Su_1996} solution at the leading edge of the wave, where our code respects causality and restricts the propagation speed of the wave to $c$; this constraint is violated in the diffusion approximation that \citeauthor{Su_1996} adopt. However, we can still compare to their solution in the region where $F_r \ll cE_r$ and diffusion is a good approximation. In this region, we obtain excellent agreement with the semi-analytic solution, as shown in \autoref{fig:marshak}. Note that the difference between our numerical solution and the ``exact'' solution at $x \gtrsim 3\times 10^{-3}$ cm is not an error in our solution. Rather, it is a result of our code properly capturing the finite speed of light, while the semi-analytic solution does not.

\subsubsection{Su-Olson problem}
\label{ssec:suolson}
We next compute a problem involving radiation penetrating a cold medium but with an internal radiation source rather than a radiation source at the boundary. This problem is defined in dimensionless units where $a_r = c = 1$, with opacities $\kappa_P = \kappa_R = 1$, a constant density $\rho = 1$, and a radiation source
\begin{align}
    S(x,t) =
    \begin{cases}
        Q \, a_r T_H^4 & 0 \leq x < x_0 \, \text{and} \, t < t_0 \, , \\
        0              & x \geq x_0 \, \text{or} \, t \geq t_0 \, ,
    \end{cases}
\end{align}
where we have a normalisation factor $Q = (2 x_0)^{-1}$, radiation source temperature $T_H = 1$, and spatial extent of the source $x_0 = 0.5$ and temporal extent $t_0 = 10$. The initial radiation and gas energies are zero in the idealized problem, but we set them to $10^{-10}$ in our simulation since the radiation solver requires nonzero gas and radiation energies. The gas velocity is zero. We adopt reflecting boundary conditions on the domain $[0, 30]$ on a grid of $N_x = 1500$ cells. We do not reduce the speed of light for this test.

\begin{figure}
    \includegraphics[width=\columnwidth]{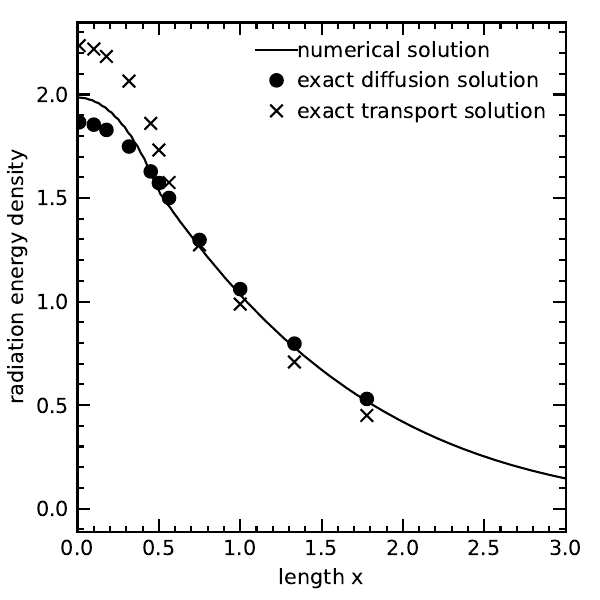}
    \caption{The Su-Olson test problem (\autoref{ssec:suolson}). The numerical solution is the solid line, with the exact diffusion solution shown as circles and the exact transport solution shown as crosses.}
    \label{fig:suolson}
\end{figure}

When using the heat capacity given by \autoref{eq:heat_capacity}, a semi-analytic solution of the angle-dependent transport equation may be obtained with a Fourier-Laplace transform \citep{Su_1997}. This solution assumes that $\vc{v} = 0$ at all times, so we drop all $v/c$ terms for this problem. We show our numerical solution using CFL number $0.4$ at time $t = 10$  in \autoref{fig:suolson}; for comparison, we also show the exact transport solution and the exact diffusion solution. We find that with the \cite{Levermore_1984} closure, we obtain a solution in between the diffusion solution and the transport solution. While it makes little difference at $t = 10$, we find better agreement with the transport solution at earlier times when using the \cite{Minerbo_1978} closure (not shown). In this problem, some regions near the internal radiation source (located at $0 \leq x < 0.5$) have Eddington factors $\chi < 1/3$, which cannot be represented by any local closure of the form given by \autoref{eq:M1_closure}.  Nonetheless, we obtain a solution that is more accurate than one would obtain by using a radiation diffusion equation.

\subsubsection{Radiation-matter energy exchange}
\label{section:equilibrium}
We next isolate the implicit matter-radiation energy exchange solver by solving a problem with no transport. Following \citet{Turner_2001}, we set up a uniform domain with periodic boundary conditions, where the gas and radiation are initially out of thermal equilibrium. The initial radiation energy density is $E_r = 10^{12} \, \text{erg} \, \text{cm}^{-3}$ and, the initial gas energy density $E_g = 10^2 \, \text{erg} \, \text{cm}^{-3}$. The density $\rho = 10^{-7} \, \text{g} \, \text{cm}^{-3}$ and the specific opacity $\kappa_P = 1.0 \, \cm^{2} \gm^{-1}$. Rather than using a constant heat capacity (as \citealt{Turner_2001} do) we use the heat capacity given by \autoref{eq:heat_capacity}, which allows us to obtain an algebraic solution for the matter temperature $T$ as a function of time $t$:
\begin{align}
    T^4 & = \left( T_{0}^4 - \frac{\hat c}{c} \tilde E_0 \right) \, \exp \left[ -\frac{4}{\alpha} \left( a_r + \frac{\hat c}{c} \frac{\alpha}{4} \right) \kappa \rho c t \right] \, + \, \frac{\hat c}{c} \tilde E_0 \, .
    \label{eq:rsla_temperature}
\end{align}
where $E_0 = E_{g} + (c/\hat c) \, E_{r}$ and $\tilde E_0 = E_0 \left[ a_r + (\hat c / c) (\alpha / 4) \right]^{-1}$ are constant as a function of time. Taking the limit $t \rightarrow \infty$, we immediately see that the equilibrium temperature $T_{\text{eq}}$ is modified whenever $\hat c \neq c$, contrary to previous claims in the literature:
\begin{align}
    T_{\text{eq}}^4 & = \frac{\hat c}{c} \tilde E_0 = \frac{\hat c}{c} E_0 \left[ a_r + \left(\frac{\hat c}{c} \right) \frac{\alpha}{4} \right]^{-1} \, .
\end{align}
Fundamentally, this occurs whenever RSLA is employed (and $\hat c \neq c$) because the quantity $E_0 = E_g + (c / \hat c) \, E_r$ is conserved in this problem, \emph{not} the total energy $E_{\text{tot}} = E_g + E_r$. This is a generic failing of the RSLA, which does not conserve total energy. However, in practice when the boundary conditions are such that the quantity $E_0$ is \emph{not} conserved, the physically correct steady-state solution may still be obtained -- this is the situation for all the radiation-hydrodynamics test problems we present in \autoref{ssec:radhydro_tests}, and is also the situation for most applications of interest.

\begin{figure}
    \includegraphics[width=\columnwidth]{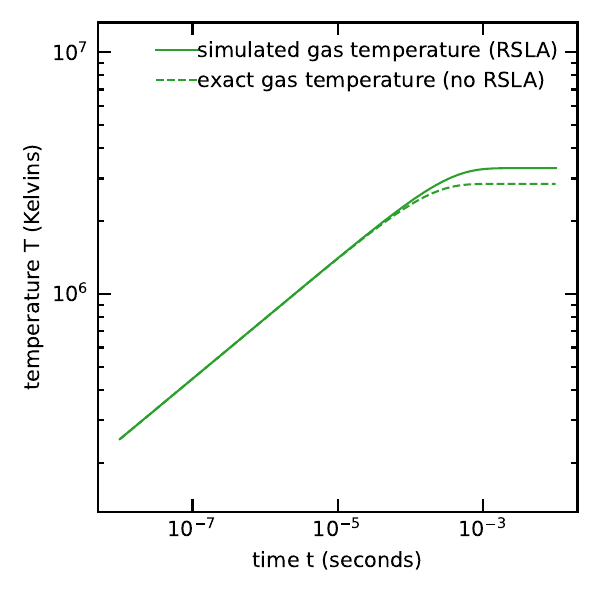}
    \caption{Time evolution of the gas temperature in the radiation-matter coupling test (\autoref{section:equilibrium}). The simulated gas temperature using RSLA is shown as the solid line. The exact solution for the physically-correct gas temperature (i.e., without RSLA; \autoref{eq:rsla_temperature}) is shown as the dashed line.}
    \label{fig:radcoupling}
\end{figure}

To test \quokka's ability to recover the analytic solution, we simulate the problem using a constant timestep $\Delta t = 10^{-8} \s$ until $t = 10^{-2} \s$. We use a reduced speed of light $\hat c = 0.1 c$. We show the time evolution of the matter temperature in \autoref{fig:radcoupling} both when using RSLA and without. We find that the numerical solution agrees with the exact solution to better than one part in $10^{5}$ at each timestep for both cases. The RSLA equilibrium temperature is approximately 20 per cent higher than the physically correct equilibrium temperature.

\subsubsection{Shadow test}
\label{section:shadow_test}
We next illustrate the performance of our radiation solver in two dimensions with a shadowing test based on that of \cite{Hayes_2003}; our test differs from theirs only in that we use planar $x-y$ geometry instead of cylindrical $r-z$ coordinates. The goal of this test is evaluate how well an RHD scheme casts sharp shadows, recovering the geometric optics limit that should prevail when the optical depth is low. This problem consists of a rectangular domain of $1.0 \cm$ by $0.12 \cm$ with a streaming radiation source incident from the left boundary and an outflow boundary condition on the right. The lower boundary is reflecting, and the upper boundary allows outflow. In the middle lies an optically-thick cylinder. The initial gas and radiation temperatures are $290 \K$ and the incident flux has a radiation temperature of $1740 \K$. The background has a density of $\rho_{\text{bg}} = 10^{-3} \gm \cm^{-3}$, and the cylinder has a density of $\rho_{\text{cl}} = 1.0 \gm \cm^{-3}$. The gas has an opacity $\rho \kappa = (\rho/\rho_{\text{bg}})^2 \, 0.1 \cm^{-1}$, a mean molecular weight $\mu = 10 \, m_{\rm H}$, and an adiabatic index $\gamma = 5/3$. We allow two levels of mesh refinement on top of a base grid of $280 \times 80$ cells, tagging cells for refinement when the relative gradient in radiation energy density exceeds $0.1$. The problem is evolved until $t = 5 \times 10^{-11} \s$ with a CFL number of $0.4$. We do not reduce the speed of light for this test. For this problem, we find it necessary to reduce the relative tolerance of the implicit matter-radiation coupling solver to $10^{-15}$. Otherwise, there are unphysical radiation shocks reflected from the cylinder.

\begin{figure*}
    \includegraphics[width=\textwidth]{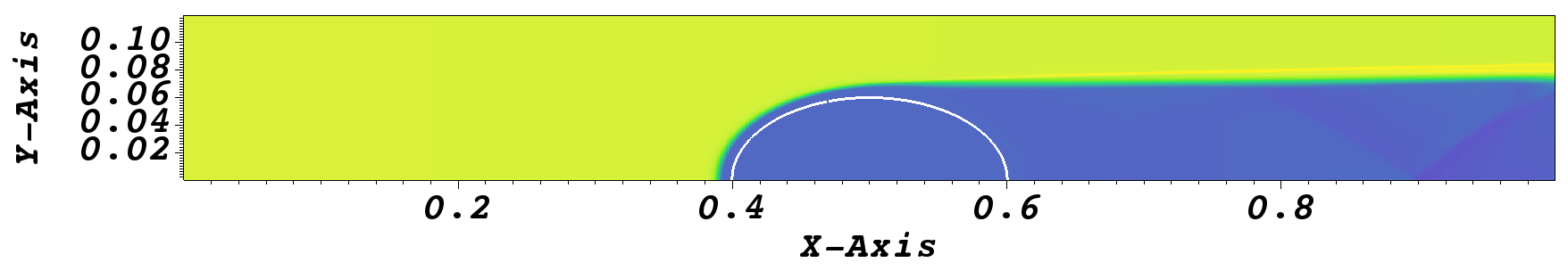}
    \caption{Results for the shadow test (\autoref{section:shadow_test}). Color shows the radiation temperature. The solid line shows a density contour indicating the position of the `cloud.'}
    \label{fig:shadow}
\end{figure*}

We show the radiation temperature at the end of the simulation in \autoref{fig:shadow}. After the beam of light has crossed the domain, we find a sharp shadow cast behind the cylinder, as one would expect physically. However, there are some residual artefacts from a transient beam of light that initially curved around the cylinder and reflected against the lower boundary, as seen in the low-temperature shock-like features within the shadow near the right edge of the domain. This appears to be an unavoidable artefact of using a local VET closure. (This does not occur when using the Eddington tensor obtained from the geometric optics limit, i.e. $f_{xx} = 1$.) Overall, this test shows that \quokka~produces qualitatively correct results for semitransparent problems.

\subsubsection{Beam test}
\label{ssec:beam}
\begin{figure}
    \includegraphics[width=\columnwidth]{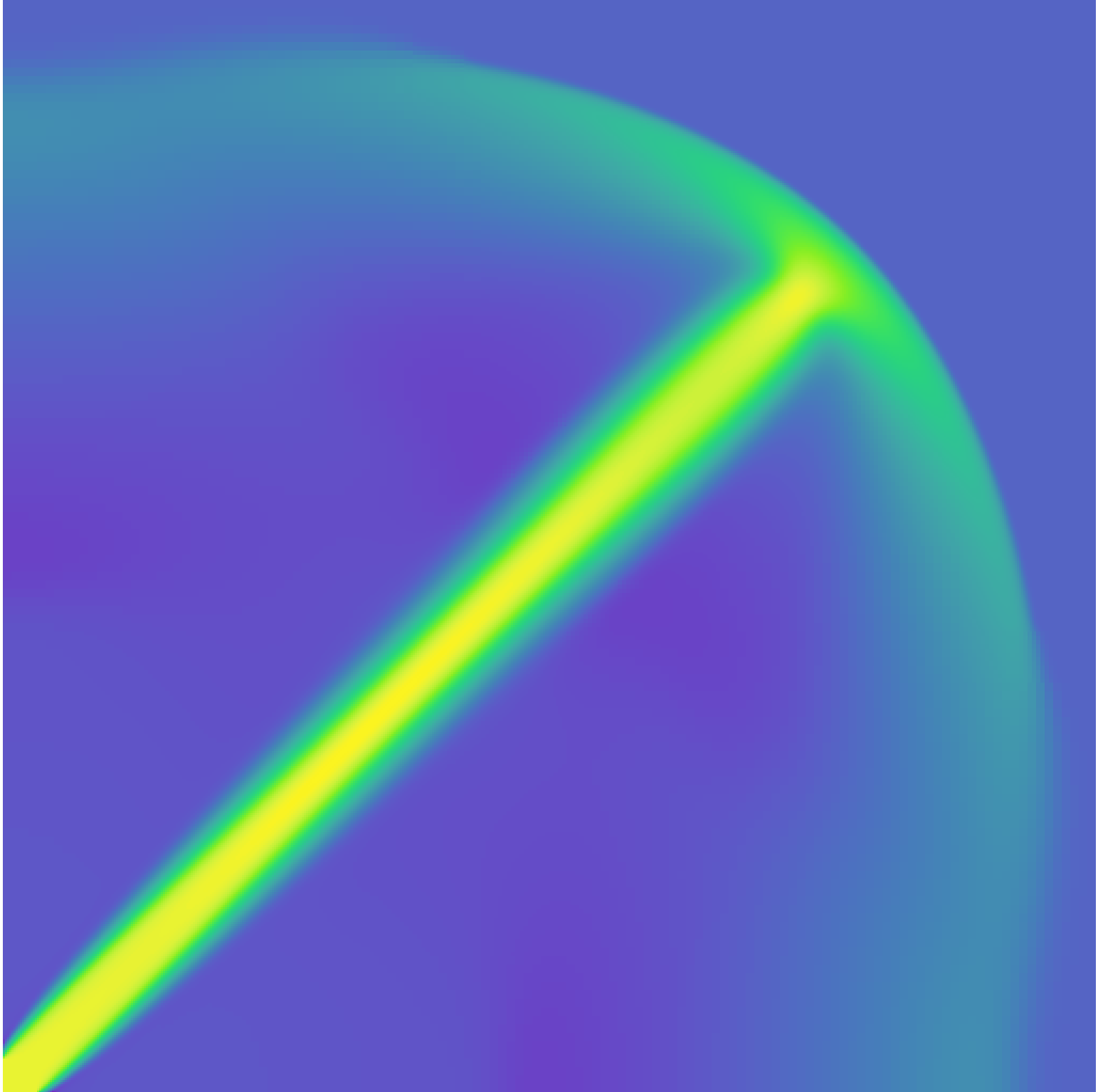}
    \caption{The beam test in vacuum (\autoref{ssec:beam}). Colour shows the logarithm of the radiation energy density.}
    \label{fig:beam}
\end{figure}

We next test our code on a beam or `searchlight' test involving streaming radiation propagating without any absorption, adopting parameters from \cite{Gonzalez_2007} with the only modification in that we move the beam to the lower left corner of the box. The domain is a square box $[0 \cm, 2 \cm]^2$ with constant density $\rho = 1.0 \gm \cm^{-3}$, gas and radiation temperature $T = 300 \K$, and zero opacity. A beam of radiation enters the domain at a $45 \deg$ angle from the lower left corner ($x < 0.0625 \cm$ or $y < 0.0625 \cm$) with a radiation temperature of $1000 \K$. We use AMR with a base grid of $128^2$ and two levels of refinement to simulate this problem, refining wherever the relative gradient of the radiation energy density exceeds $0.1$. For this problem, we use PLM reconstruction for the radiation variables in order to avoid oscillations near the leading edge of the beam. We use a CFL number of $0.4$. We show the radiation energy density at time $t =  1.172 \, (L/c)$, where $L$ is the box size, in \autoref{fig:beam}. The beam stays relatively narrow as it crosses the box, but at the leading edge of the beam, we see there is a transient bow shock-like feature which is due to our use of a local VET closure. Our code performs reasonably well on this problem, showing only a small amount of diffusion of the beam as it propagates. The bow shock feature appears to be an unphysical `radiation shock' that can occur due to the nonlinear behavior of non-constant local VET closures. For instance, it can be shown that the radiation moment equations without source terms with the \cite{Levermore_1984} closure are mathematically identical to the hydrodynamic equations of an ultrarelativistic gas \citep{Hanawa_2014}. After the leading edge of the beam has crossed the box, the bow shock feature leaves the box and a narrow beam of light remains.

\subsection{Radiation hydrodynamics}
\label{ssec:radhydro_tests}

Our final suite of tests use the full suite of physics in \quokka, and involve coupled radiation and hydrodynamics.

\subsubsection{Radiation pressure tube}
\label{section:radtube}
Our first radiation-hydrodynamic test is the the radiation pressure tube problem of \cite{Krumholz_2007}. This problem is designed to show that the radiation pressure gradient can stably balance the gas pressure gradient both in the regime where radiation pressure dominates and the in the regime where gas pressure dominates for a problem where the optical depth is sufficiently large that the radiation is in the equilibrium diffusion regime. We adopt the opacities $\kappa_P = \kappa_R = 100 \cm^2 \gm^{-1}$, mean molecular weight $\mu = 2.33 \, m_H$, and adiabatic index $\gamma = 5/3$. The exact steady-state solution in the diffusion approximation is given by the solution to the differential equations
\begin{align}
    \frac{d \rho}{dx}  & = -\frac{\mu}{k_B T} \left( \frac{k_B}{\mu} \rho + \frac{4}{3} a_r T^3 \right) \frac{dT}{dx} \, , \\
    \frac{d^2 T}{dx^2} & = -\frac{3}{T} \left(\frac{dT}{dx}\right)^2 + \frac{1}{\rho} \frac{d\rho}{dx} \frac{dT}{dx} \, ,
\end{align}
where the left-side temperature, density, and density gradient are $T_0 = 2.75 \times 10^7 \K$, $\rho_0 = 1.0 \gm \cm^{-3}$, and ${d\rho_0}/{dx} = 0.005 \gm \cm^{-4}$. We solve this equation on the domain $[0 \cm, 128 \cm]$ in order to obtain the initial conditions for this problem. The left and right side initial conditions are adopted as Dirichlet boundary conditions for our simulation. The reduced speed of light $\hat c$ is set to $10 \, c_{s,0} \approx 4.03 \times 10^8 \cm \s^{-1}$, where $c_{s,0}$ is the sound speed at the left boundary.

\begin{figure}
    \includegraphics[width=\columnwidth]{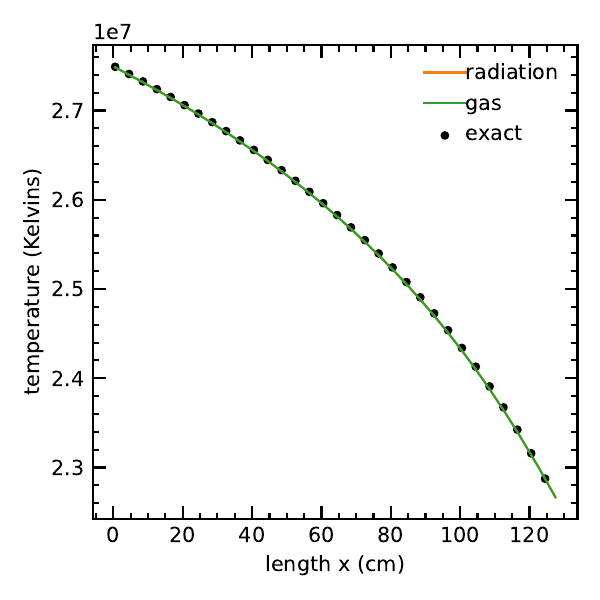}
    \caption{Temperature profiles for the radiation pressure tube test (\autoref{section:radtube}). The radiation and gas temperatures are both shown, but only the latter line is visible because the two temperatures are nearly identical. The temperature for the exact diffusion solution is shown in the black circles. The simulated and exact temperatures agree to within 0.2 per cent.}
    \label{fig:radiation_pressure_tube}
\end{figure}

After evolving for a sound crossing time $t = L_x / c_{s,0} \approx 3.177 \times 10^{-6} \s$ with a CFL number of $0.4$ on a grid of 128 cells, we obtain the numerical solution shown in \autoref{fig:radiation_pressure_tube}. We note that these parameters imply an optical depth per cell of $\tau_{\text{cell}} \sim 10^2$, so this problem tests the accuracy of our numerical methods in the asymptotic diffusion regime where $\tau_{\text{cell}} \gg 1$. Our numerical solution agrees with the initial conditions (obtained from the exact diffusion solution) to better than $0.2$ per cent. Since the boundary conditions do not require conservation of the quantity $E_0$ (see \autoref{section:equilibrium}), we find that we are able to obtain the physically correct solution even when $\hat c \neq c$.

\subsubsection{Optically-thin radiation-driven wind}
\label{section:thinwind}
In order to test the radiation-gas momentum coupling in the optically-thin limit, we next simulate a radiation-driven wind in the limit of very low optical depth. We consider an isothermal gas with sound speed $c_T = 0.2 \km \s^{-1}$, with constant opacities $\kappa_P = 0$ and $\kappa_R = 5 \cm^2 \gm^{-1}$. A flux of radiation $F_{r,0}$ enters the computational domain from the left side at $x=0$, inducing an acceleration $a_0 = \kappa_R F_{r,0}/c$ in the gas; we choose the density of the gas low enough that the optical depth is negligible, so the flux and acceleration are constant across the domain. Consider a fluid parcel moving at Mach number $\mathcal{M}_0$ at $x = 0$, where the radiation flux enters the domain. Integrating the gas momentum equation with respect to position $x$ yields a Bernoulli equation for the Mach number $\mathcal{M}$ as a function of position
\begin{align}
    \frac{1}{2} \mathcal{M}_0^2 = \frac{1}{2} \mathcal{M}^2 + \log \left({\frac{\mathcal{M}_0}{\mathcal{M}}}\right) - \left( \frac{x}{L} \right) \, ,
    \label{eq:wind}
\end{align}
where $L = c_T^2/a_0$ is the characteristic acceleration length of the problem. If the density at $x = 0$ is $\rho_0$, then from conservation of mass, in steady state the density as a function of position is $\rho = (\mathcal{M}_0/\mathcal{M}) \rho_0$.\footnote{We neglect gravitational forces in this problem, but we note that our solution is formally equivalent to the Eddington ratio $\eta_{\text{Edd}} = 2$ case of the plane-parallel radiation-inhibited Bondi accretion problem of \cite{Skinner_2013}.} For our test we choose $\mathcal{M}_0 = 1.1$, and we set $\rho_0 = 3.897212 \times 10^{-19} \gm \cm^{-3}$; for our chosen value of $\kappa_R$, this yields an optical depth $\tau = 10^{-6}$ from $x=0$ to $x=L$.

\begin{figure}
    \includegraphics[width=\columnwidth]{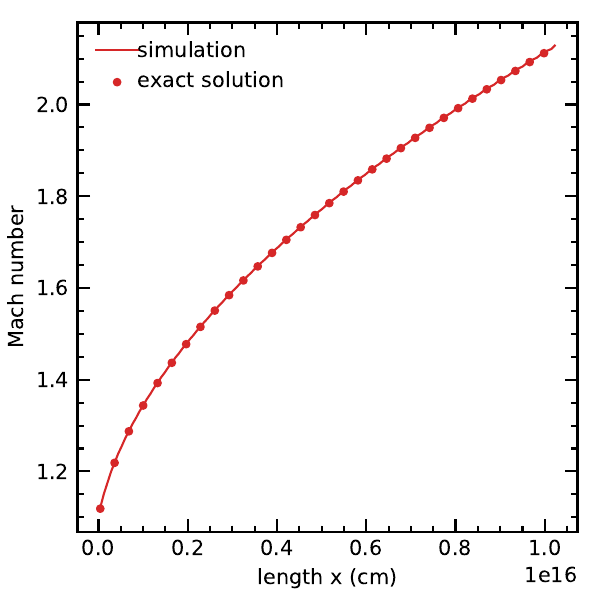}
    \caption{Mach number as a function of position for a radiation-driven wind (\autoref{section:thinwind}). The solid line is the \quokka~numerical solution, while the points show the exact solution given by \autoref{eq:wind}.}
    \label{fig:wind}
\end{figure}

To simulate this problem, we set up a domain from $x = 0$ to $x=L$, resolved by $N_x = 128$ cells. We initialize the system with the exact solution (\autoref{eq:wind}), and also use the exact solution to impose Dirichlet boundary conditions on the density, velocity, and radiation flux. We evolve the system until $t=10\,(L/c_T)$ using a CFL number of 0.4; for this calculation we use an isothermal Riemann solver. We set the reduced speed of light to $\hat c = 10 \mathcal{M}_1 c_T$, where $\mathcal{M}_1$ is the Mach number at the right-side boundary. In \autoref{fig:wind}, we show the exact solution for the Mach number (circles) compared to the solution produced by \textsc{Quokka} (solid line), finding excellent agreement.

\subsubsection{Subcritical radiative shock}
\label{section:radshock}
\begin{figure}
    \includegraphics[width=\columnwidth]{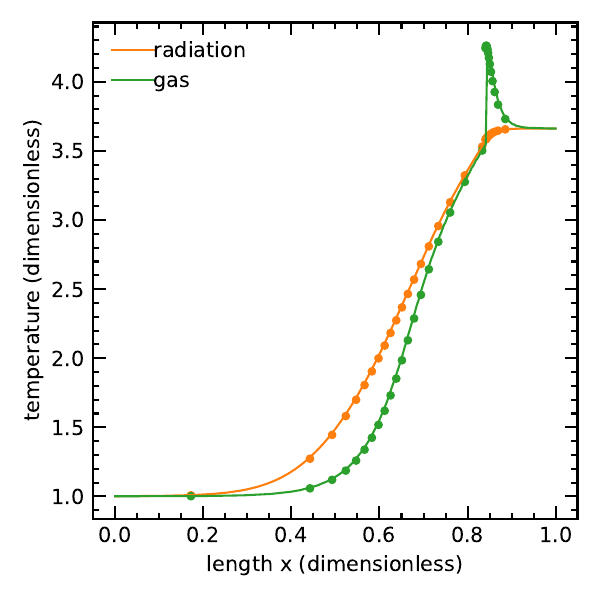}
    \caption{Radiation and matter temperatures in a subcritical radiative shock with $\mathcal{M} = 3$ (\autoref{section:radshock}). The simulation results as shown as solid lines, while the exact steady-state solution is shown as circles.}
    \label{fig:radshock}
\end{figure}

We next simulate a subcritical radiative shock, following the set-up used by \cite{Skinner_2019} with the dimensionless parameters for the Mach $\mathcal{M} = 3$ example given by \cite{Lowrie_2008}. We scale to cgs units with the opacities $\kappa_P = \kappa_R = 577 \cm^{2} \gm^{-1} (1 \gm \cm^{-3} / \rho)$, mean molecular weight $\mu = m_{\rm H}$, and adiabatic index $\gamma = 5/3$. The left-side state consists of $\rho_L = 5.69 \gm \cm^{-3}$, velocity $v_L = 5.19 \times 10^7 \cm \s^{-1}$, and temperature (gas and radiation) $T_L = 2.18 \times 10^6 \Kelvin$. The right-side state is $\rho_R = 17.1 \gm \cm^{-3}$, $v_R = 1.73 \times 10^7 \cm \s^{-1}$, and $T_R = 7.98 \times 10^6 \K$. These states are also used as Dirichlet boundary conditions for the simulation. In order to exactly match the assumptions used in the semi-analytic solution of \cite{Lowrie_2008}, we use the Eddington approximation (i.e., $\textsf{P}_r = (1/3) E_r \textsf{I}$) to close the radiation pressure tensor for this problem.\footnote{We provide a \textsc{Python} code that computes the semi-analytic solution for radiative shocks using the Eddington approximation \citep{Lowrie_2008} in our \faGithub\href{https://github.com/BenWibking/quokka-code}{GitHub repository}.} Following \cite{Skinner_2019}, we use a reduced speed of light $\hat c = 10(v_L + c_{s,L})$, where $c_{s,L}$ is the adiabatic sound speed of the left-side state.  We use a CFL number of 0.4 and evolve until $t = 10^{-9} \s$ on a grid of $512$ cells on the domain $[0 \cm, 0.01575 \cm]$, with the discontinuity placed between the left- and right-side states $x_0 = 0.0130 \cm$. The shock drifts $1.5$ per cent of the domain length to the right from the location of the initial discontinuity, which may be due to a combination of the initial numerical transient and our use of the asymptotic states as boundary conditions, rather than the exact states expected at a finite distance from the shock location. This makes the steady-state location of the shock on the simulation grid not well-defined. After accounting for this drift, the agreement between the numerical and semi-analytic solution is excellent, as shown in \autoref{fig:radshock}. We find the that the relative error of the gas temperature in $L_1$ norm is $0.4$ per cent, which is at least as good as the solution of \cite{Skinner_2019} for the same spatial resolution.  In this problem, we find that using shock flattening is essential to obtain a non-oscillatory temperature structure for the Zel'dovich spike (the gas temperature discontinuity shown in \autoref{fig:radcoupling}; \citealt{Zeldovich_1967}).

\subsubsection{Radiation-driven dust shell}
\label{section:shell}
As a final example, we consider a non-steady-state radiation hydrodynamics problem: the radiation-driven dust shell problem from \citet[hereafter \citetalias{Skinner_2013}]{Skinner_2013}, consisting of an initial shell of dusty gas placed at radius $r_0$ with the radial density profile:
\begin{align}
    \rho(r) = \frac{M_\text{sh}}{4\pi r^2 \sqrt{2\pi \sigma_{\text{sh}}^2}} \exp \left( -\frac{(r - r_0)^2}{2\sigma_{\text{sh}}^2} \right) \, ,
    \label{eq:shell_density}
\end{align}
where $M_{\text{sh}}$ is the mass of the shell and $\sigma_{\text{sh}}$ is the thickness of the shell. We place a point-like source of radiation, representing a central star, at $r = 0$. The radiation source is smoothed so that it can be resolved on the computational grid, using a Gaussian profile of the form
\begin{align}
    j(r) = \frac{L_{\star}}{(2\pi \sigma_{\star}^2)^{3/2}} \exp \left( -\frac{r^2}{2 \sigma_{\star}^2} \right)
\end{align}
where $L_{\star}$ is the luminosity of the source and $\sigma_{\star}$ is a smoothing parameter defining the spatial extent of the source. Under the thin-shell approximation and neglecting gas pressure forces, \citetalias{Skinner_2013} obtain an equation of motion for the shell. Starting from rest, the resulting shell velocity, written in terms of the shell Mach number $\mathcal{M}_{\text{sh}}$, is
\begin{align}
    \mathcal{M}_{\text{sh}} \equiv \frac{dR}{dT} = \sqrt{2} \mathcal{M}_0 \sqrt{1 - \frac{1}{R}} \, .
\end{align}
with dimensionless radius $R \equiv r/r_0$, dimensionless time $T \equiv t/t_0$, characteristic time $t_0 = r_0/c_T$, reference sound speed $c_s$, and reference Mach number $\mathcal{M}_0$:
\begin{align}
    \mathcal{M}_0 = c_s^{-1} \sqrt{\frac{L_{\star} \kappa_R}{4\pi r_0 c}} \, .
\end{align}
Following the parameters used by \citetalias{Skinner_2013}, we set $\kappa_P = \kappa_R = 20 \cm^2 \gm^{-1}$, $c_s = 2 \km \s^{-1}$, $r_0 = 5 \pc$, $M_{\text{sh}} = 5 \times 10^5$ \Msun, and $L_{\star} = 2 \times 10^{42} \erg \s^{-1}$. We note that $c_s$ is only a reference sound speed and does not change the thin-shell solution since pressure forces are assumed to be negligible. We adopt values of $\sigma_{\star} = 0.3 r_0$ and $\sigma_{\text{sh}} = 0.3 r_0/(2\sqrt{2 \log 2})$.

\begin{figure}
    \includegraphics[width=\columnwidth]{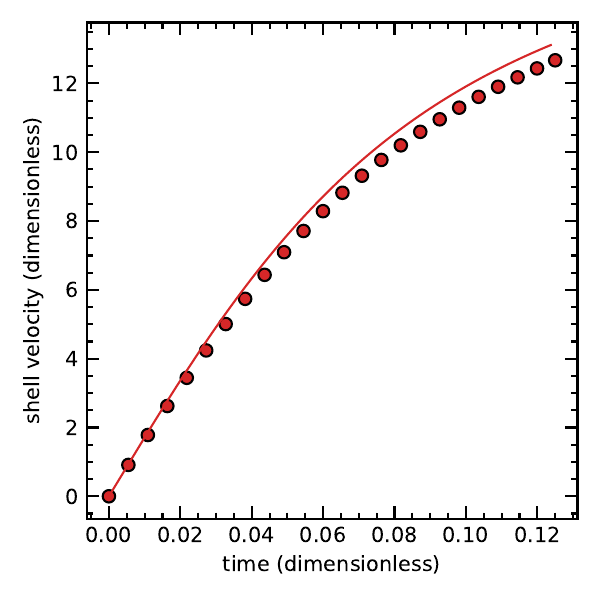}
    \caption{Shell velocity as a function of dimensionless time $T = t/t_0$ in the radiation-driven dust shell test (\autoref{section:shell}). The exact thin-shell solution is shown as the solid line, while the circles show the mass-weighted shell velocity from the simulation.}
    \label{fig:radshell}
\end{figure}

We initialize our simulation of this problem using the density profile (\autoref{eq:shell_density}) and the quasi-static radiation energy and flux derived by \citetalias{Skinner_2013}. We initialize the gas temperature in equilibrium with the radiation temperature. A density floor is set at $\rho_{\text{floor}} = 10^{-8} \rho_0$, where $\rho_0 = M_{\text{sh}} / (4 \pi r_0^3 / 3)$. A pressure floor is likewise set at $P_{\text{floor}} = 10^{-8} P_0$, where $P_0 = \gamma \rho_0 c_s^2$. We use an adiabatic equation of state with $\gamma = 5/3$ and a mean molecular weight $\mu = 2.33 m_H$, where $m_H$ is the mass of the hydrogen atom. We use a uniform grid of $128^3$ and PLM reconstruction for both hydrodynamic and radiation variables. Following \citetalias{Skinner_2013}, we reduce the speed of light to $\hat c = 860 \, c_s$. The simulation is evolved until $t = 0.125 \, t_0$ using a CFL number of $0.3$. The shell velocity as a function of time is shown in \autoref{fig:radshell}. Our simulation has values slightly lower than expected from the thin-shell solution, whereas \citetalias{Skinner_2013} find simulated shell velocities slightly higher than the thin shell solution. Exact agreement cannot be expected since \citetalias{Skinner_2013} do not specify their values for the parameters $\sigma_{\star}$ and $\sigma_{\text{sh}}$ and differences in implementation details of our radiation hydrodynamic solvers. Overall, we find very good agreement between the simulation (shown as crosses) and the thin-shell solution (solid line).

\section{Performance and scaling}
The entire motivation for \quokka~is to achieve high performance on RHD problems run on GPUs. We therefore next test the performance and scaling of the code. All the tests we present were performed on the Gadi supercomputer at the National Computational Infrastructure\footnote{\url{https://nci.org.au/our-systems/hpc-systems}}, using the gpuvolta nodes. Each node has 2 24-core Intel Xeon Platinum 8268 (Cascade Lake) 2.9 GHz CPUs and 4 Nvidia Tesla Volta V100-SXM2-32GB GPUs connected to each other in an all-to-all topology with NVLink 2.0. Nodes are coupled via HDR InfiniBand in a Dragonfly+ topology.

\label{section:performance}
\subsection{Weak scaling}
We first demonstrate that \textsc{Quokka} has excellent parallel scaling efficiency when keeping the number of computational cells fixed per GPU (referred to as \emph{weak scaling}). For our first test of weak scaling, we show the scalability of the hydrodynamics solver on uniform grids, disabling mesh refinement and radiation. We simulate a Sedov-Taylor blast wave \citep{Sedov_1959,Taylor_1946} in a 3D periodic box on the domain $[-1, 1]$ in each coordinate direction. The initial conditions consist of a spherical region of high pressure $P = 10$ for radii $r < 0.1$ and low pressure $P = 0.1$ for $r \ge 0.1$, with a uniform density of $\rho = 1$ and zero velocity, for an ideal gas with adiabatic index $\gamma = 5/3$.

We run with a varying number of GPUs with two $256^3$ grids per GPU, increasing the resolution of our simulation as we extend to greater numbers of GPUs. However, a power-of-two resolution increase does not easily map onto a jump from one GPU to four GPUs, so the single-GPU simulation only uses a grid size of $256^3$. The grid size of the simulations therefore ranges from $256^3$ (for 1 GPU) to $2048^3$ (for 256 GPUs). We set the \textsc{AMReX} domain decomposition parameters \texttt{blocking\_factor} and \texttt{max\_grid\_size} to a value of $128$, leading the computational grid to be decomposed into arrays of size $128^3$. (We also tested local grid sizes of $256^3$ but found only a few per cent performance improvement on this problem.) We use one MPI rank per GPU for all simulations. The CFL number is $0.25$ and we evolve for $100$ timesteps for each simulation. We assess performance by counting the total number of cell-updates and dividing by the number of GPUs in order to obtain the performance figure-of-merit in the units of 1 million cells (or zones) per timestep per GPU per second (Mzones/GPU/s).  We report the results in \autoref{table:weak_hydro_scaling}.

We find a $\approx 40\%$ drop in performance per GPU when going from 1 GPU to 4 GPUs, corresponding to using all 4 GPUs on a single node of the compute cluster. We hypothesize that this is due to the limited communication bandwidth between GPUs on a node. For intra-node scaling on CPUs, \citet{Stone_2020} report a similar decrease in performance when going from one CPU to all the CPUs on a node for \textsc{Athena++}, which they attribute to limitations of memory bandwidth. However, significantly different scaling behavior is observed when running the \textsc{K-Athena} hydrodynamics code on GPUs \citep{Grete_2019} on the Summit supercomputer\footnote{\url{https://www.olcf.ornl.gov/olcf-resources/compute-systems/summit/}}, finding a $99$ per cent weak scaling efficiency going from 1 GPU to 6 GPUs on a single node, so there may be some inefficiency in our current GPU-to-GPU communication method. We find that using CUDA-aware MPI does not improve performance for our code. However, we observe only a modest drop in performance per GPU when going from 1 node (4 GPUs) to 64 nodes (256 GPUs), yielding a parallel efficiency of 83 per cent on 64 nodes when compared to running on 1 node.  We could not run on larger numbers of GPU nodes due to job size limitations, but we expect scaling to continue to thousands of GPUs based on the parallel scaling observed for other GPU hydrodynamics codes based on \textsc{AMReX}, such as \textsc{Castro} \citep{Almgren_2020}.

In \autoref{table:weak_hydro_scaling_plm}, we show the same performance numbers as in \autoref{table:weak_hydro_scaling}, but using PLM reconstruction for each simulation instead of PPM reconstruction. We find that the performance improves significantly on a single GPU, going from $113$ million zone-updates per second to $158$ million zone-updates per second. However, communication overheads limit the relative performance improvement when using large numbers of nodes, as the $64$-node case goes from $59$ million zone-updates per GPU per second using PPM to only $65$ million zone-updates per GPU per second using PLM. Since the computations on each local grid are less expensive with PLM but the communication costs remain the same, the scaling efficiency decreases slightly as well, from $83$ per cent to $76$ per cent.

\begin{table}
    \begin{tabular}{l|r|r|r|r}\hline
        Nodes   & GPUs  & Mzones/GPU/s  & Scaling efficiency (\%) & Grid size                         \\\hline
        \begin{filecontents*}{weak_scaling_hydro.csv}
nodes,gpus,mzones,mzones_per_gpu,mzones_per_gpu_ideal,GPU_FillBoundary,GPU_scaling,GPU_scaling_1node,size
0.25,1,113.34,113.34,116.15,2.42,100.00,--,256^3
1,4,282.72,70.68,115.68,38.90,62.36,100.00,512^3
8,32,1847.98,57.75,114.04,49.36,50.95,81.70,1024^3
64,256,14975.34,58.50,113.83,48.61,51.61,82.76,2048^3
\end{filecontents*}
\csvreader
        {weak_scaling_hydro.csv}{1=\nodes,2=\gpus,3=\mzones,4=\mzonespergpu,5=\mzonespergpuideal,6=\gpufill,7=\scaling,8=\scalingnode,9=\size}
        {\nodes & \gpus & \mzonespergpu & \scalingnode       & $\size$ \\}
    \end{tabular}
    \caption{Weak scaling efficiency for hydrodynamics with PPM reconstruction as a function of the number of GPUs for a Sedov blast wave with periodic boundary conditions.}
    \label{table:weak_hydro_scaling}
\end{table}

\begin{table}
    \begin{tabular}{l|r|r|r|r}\hline
        Nodes   & GPUs  & Mzones/GPU/s  & Scaling efficiency (\%) & Grid size                         \\\hline
        \begin{filecontents*}{weak_scaling_hydro_plm.csv}
nodes,gpus,mzones,mzones_per_gpu,GPU_scaling,GPU_scaling_1node,size
0.25,1,157.75,157.75,100.00,--,256^3
1,4,343.77,85.94,54.48,100.00,512^3
8,32,2392.59,74.77,47.40,87.00,1024^3
64,256,16684.81,65.18,41.32,75.84,2048^3
\end{filecontents*}
\csvreader
        {weak_scaling_hydro_plm.csv}{1=\nodes,2=\gpus,3=\mzones,4=\mzonespergpu,5=\scaling,6=\scalingnode,7=\size}
        {\nodes & \gpus & \mzonespergpu & \scalingnode       & $\size$ \\}
    \end{tabular}
    \caption{Weak scaling efficiency for hydrodynamics with PLM reconstruction as a function of the number of GPUs for a Sedov blast wave with periodic boundary conditions.}
    \label{table:weak_hydro_scaling_plm}
\end{table}

We next test the scaling behaviour for full radiation hydrodynamics solver on uniform grids. \autoref{table:weak_radhydro_scaling} lists the performance per GPU and parallel efficiency measured with respect to single-node performance for the radiation-driven shell test problem run for $50$ timesteps. Since we have many radiation substeps per hydrodynamic step (set here to 10; see \autoref{sssec:sync}), the performance metric in units of Mzones/GPU/s is lower by a factor comparable to but somewhat smaller than the number of radiation substeps per hydro step; a single radiation update is slightly less costly than a single hydrodynamic update. In this case, we observe a steeper drop in performance when going from 1 GPU to 4 GPUs (approximately a factor of 2). The lower parallel efficiency is not surprising, since \emph{each} radiation substep requires communicating boundary conditions between grids, so the amount of inter-GPU communication per hydro timestep increases significantly for radiation hydrodynamics. Nonetheless, as is the case for hydrodynamics, there is little additional performance penalty when scaling from 1 node to 64 nodes. We measure a parallel efficiency in this case of $76$ per cent.

in \autoref{table:scaling_a100}, we list the performance metrics of the code on both the Sedov problem and the radiation-driven shell problem running on a compute node with newer NVIDIA A100 GPUs. Since we only have access to a limited number of these GPUs, we only show performance data for a single GPU and a single compute node (4 GPUs). The single GPU case achieves $254$ million hydrodynamic zone-updates per second using PPM, making Quokka, as far as we are aware, the fastest PPM hydrodynamics code that currently exists. On the radiation-driven shell problem, the code achieves $39$ million radiation hydrodynamic zone-updates per second. In both cases, the performance per GPU drops by a factor of approximately 2 when using all 4 GPUs on the node. This is almost entirely due to the time spent communicating boundary conditions, as shown in the table (`B.C. fill time', which denotes the percentage of total wall time spent filling ghost cells for each local grid). If communication of boundary data and computation over the local grids could be perfectly overlapped, the parallel efficiency going from 1 GPU to 4 GPUs would be $> 99$ per cent for hydrodynamics and $> 96$ for radiation hydrodynamics.

Finally, we point out that absolute speed of \quokka~is excellent. Comparison between CPU and GPU codes is non-trivial, since it obviously depends on the CPU-to-GPU ratio on a particular compute platform. However, it is worth pointing out that \quokka's update rate per core (normalised by the number of CPU cores per compute node) for \textit{radiation}-hydrodynamics on GPU is comparable to or better than \textsc{Athena++}'s for \textit{hydrodynamics} on CPU.

\begin{table}
    \begin{tabular}{l|r|r|r|r|r|r}\hline
        Nodes   & GPUs  & Mzones/GPU/s  & Scaling efficiency (\%) & Grid size                         \\\hline
        \begin{filecontents*}{weak_scaling_radhydro.csv}
nodes,gpus,mzones,mzones_per_gpu,mzones_per_gpu_ideal,GPU_FillBoundary,GPU_scaling,GPU_scaling_1node,size
0.25,1,22.55,22.55,23.75,5.05,100.00,--,256^3
1,4,41.29,10.32,23.07,55.26,45.78,100.00,512^3
8,32,253.53,7.92,22.73,65.15,35.14,76.75,1024^3
64,256,2013.45,7.87,22.93,65.70,34.88,76.19,2048^3
\end{filecontents*}
\csvreader
        {weak_scaling_radhydro.csv}{1=\nodes,2=\gpus,3=\mzones,4=\mzonespergpu,5=\mzonespergpuideal,6=\gpufill,7=\scaling,8=\scalingnode,9=\size}
        {\nodes & \gpus & \mzonespergpu & \scalingnode       & $\size$ \\}
    \end{tabular}
    \caption{Weak scaling efficiency for radiation hydrodynamics as a function of the number of GPUs for the radiation-driven shell test (\autoref{section:shell}) with periodic boundary conditions.}
    \label{table:weak_radhydro_scaling}
\end{table}

\begin{table}
    \begin{tabular}{r|r|r|c|c}\hline
        GPUs  & Mzones/GPU/s  & B.C. fill time (\%) & Grid size & Problem                         \\\hline
        \begin{filecontents*}{performance_a100.csv}
nodes,gpus,mzones,mzones_per_gpu,mzones_per_gpu_ideal,GPU_FillBoundary,GPU_scaling,node_scaling,size,problem_type
0.25,1,254.05,254.05,259.23,2.00,100.00,--,256^3,Sedov
1,4,601.05,150.26,259.21,42.03,59.15,100.00,512^3,Sedov
0.25,1,39.04,39.04,40.30,3.12,100.00,--,256^3,Rad. shell
1,4,72.67,18.17,38.88,53.27,46.54,100.00,512^3,Rad. shell
\end{filecontents*}
\csvreader
        {performance_a100.csv}{1=\nodes,2=\gpus,3=\mzones,4=\mzonespergpu,5=\mzonespergpuideal,6=\gpufill,7=\scaling,8=\scalingnode,9=\size,10=\problemtype}
        {\gpus & \mzonespergpu & \gpufill & $\size$       & \problemtype \\}
    \end{tabular}
    \caption{Performance for both Sedov and radiating shell tests as a function of the number of GPUs for a single node with 4 NVIDIA A100 GPUs.}
    \label{table:scaling_a100}
\end{table}

\subsection{Strong scaling with AMR}
Many applications of interest will seek to minimize either the total runtime of the simulation or the total node-hours used for a simulation for a problem of a fixed size. Additionally, most applications we are interested in will benefit from or require the use of AMR.  We therefore test the ability of \textsc{Quokka} to scale an AMR radiation hydrodynamic simulation of fixed size to larger numbers of GPUs in order to either minimize total runtime or total node-hours (referred to as \emph{strong scaling}). For this test, we initialize the radiation-driven shell problem (\autoref{section:shell}) on a base grid of $256^3$ cells with two levels of mesh refinement based on the relative gradient in the gas density. We run each simulation for 50 timesteps, with a CFL number of $0.3$ and PLM reconstruction for both hydrodynamics and radiation. We set the \textsc{AMReX} domain decomposition parameters \texttt{blocking\_factor} set to 32 and \texttt{max\_grid\_size} set to a value of $128$, so that all grids are between $32^3$ and $128^3$ in size, with possible non-cubic grids at intermediate sizes. The number of GPUs used for each simulation is varied, scaling from 1 node (4 GPUs) to 8 nodes (32 GPUs). This is a particularly stringest test, since the level-by-level AMR timestepping requires that each level be computed separately, limiting the amount of parallelism that can be distributed across GPUs. There is also additional communication overhead when AMR is enabled compared to a single-level uniform grid simulation. We show the scaling results in \autoref{table:strong_scaling}. Comparing \autoref{table:weak_radhydro_scaling} and \autoref{table:strong_scaling}, the performance per GPU for a single node is lower than that of a uniform grid simulation by $\approx 50$ per cent. (A similar, although somewhat smaller, overhead when enabling AMR is also observed with CPU codes, e.g., \textsc{Athena++}; \citealt{Stone_2020}). The scaling efficiency is reasonable for 2 and 4 nodes (66 per cent for 4 nodes), but drops significantly at 8 nodes to 53 per cent parallel efficiency. We hypothesize that this is due to the small number of cell-updates per GPU once 32 GPUs are in use for this problem (approximately $211^3 \approx 9.4 \times 10^6$ cells/GPU). We find that performance on a single GPU is significantly diminished for uniform-grid problems smaller than $256^3$, so this performance drop may be largely due to the inability to use all GPU hardware threads when the amount of work per GPU is small. Similar GPU performance behavior is observed when running \textsc{K-Athena} on GPUs for varying problem sizes per GPU \citep{Grete_2019}. This effect is also magnified by the sequential nature of the level-by-level timestepping. For level $l=1$, the number of cells per GPU drops below $256^3$ for 8 GPUs, and for level $l=2$, it drops below $256^3$ for 16 GPUs. High scaling efficiency is obtained before reaching these thresholds, so it appears that reasonable performance on GPUs may be obtained with AMR when all refinement levels have at least $256^3$ cells per GPU on average. In general, obtaining the best possible GPU performance may require an adjustment to the mesh refinement parameters usually used when running on CPUs. For self-gravitating problems, scaling may be aided by the self-similar nature of gravitational collapse, leading to an approximately equal number of cells on each refinement level for appropriate refinement criteria (see discussion in \citealt{Stone_2020}).

\begin{table}
    \begin{tabular}{l|r|c|r|r|r|r}\hline
        Nodes   & GPUs  & Mzones/GPU/s  & $\left<\frac{\text{Cells}}{\text{GPU}}\right>$ & \begin{tabular}{@{}r@{}}Scaling \\ efficiency\end{tabular} & Speedup                               \\\hline
        \begin{filecontents*}{strong_scaling.csv}
nodes,gpus,mzones,mzones_per_gpu,mzones_per_gpu_ideal,GPU_FillBoundary,GPU_scaling,average_cells_per_gpu,speedup,cells_per_gpu_0,cells_per_gpu_1,cells_per_gpu_2
1,4,19.82,4.95,16.70,70.33,100.00,421.8,1.00,161.3,259.1,376.7
2,8,34.36,4.30,16.92,74.61,86.70,334.8,1.73,128.0,205.6,299.0
4,16,52.19,3.26,16.39,80.10,65.83,265.7,2.63,101.6,163.2,237.3
8,32,83.53,2.61,16.78,84.44,52.69,210.9,4.21,80.6,129.5,188.4
\end{filecontents*}
\csvreader
        {strong_scaling.csv}{1=\nodes,2=\gpus,3=\mzones,4=\mzonespergpu,5=\mzonespergpuideal,6=\gpufill,7=\scaling,8=\cellspergpu,9=\speedup,10=\cellszero,11=\cellsone,12=\cellstwo}
        {\nodes & \gpus & \mzonespergpu & ${\cellspergpu}^3$                             & \scaling                   & {\speedup}x \\}
    \end{tabular}
    \begin{tabular}{l|r|r|r|r}\hline
        Nodes   & GPUs  & $\left<\frac{\text{Cells}}{\text{GPU}}\right>_{l=0}$ & $\left<\frac{\text{Cells}}{\text{GPU}}\right>_{l=1}$ & $\left<\frac{\text{Cells}}{\text{GPU}}\right>_{l=2}$ \\\hline
        \csvreader
        {strong_scaling.csv}{1=\nodes,2=\gpus,3=\mzones,4=\mzonespergpu,5=\mzonespergpuideal,6=\gpufill,7=\scaling,8=\cellspergpu,9=\speedup,10=\cellszero,11=\cellsone,12=\cellstwo}
        {\nodes & \gpus & ${\cellszero}^3$                                     & ${\cellsone}^3$                                      & ${\cellstwo}^3$ \\}
    \end{tabular}
    \caption{Strong scaling efficiency for radiation hydrodynamics as a function of the number of GPUs for the radiation-driven shell test (\autoref{section:shell}) with periodic boundary conditions on a base grid of $256^3$ cells and 2 levels of refinement. The number of cells per GPU is computed as an average over all timesteps.}
    \label{table:strong_scaling}
\end{table}

\section{Discussion and Conclusions}
\label{section:discussion}

We conclude by discussing some of the limitations of \quokka~as it currently exists, and our plans for future expansions of the code that will address at least some of these.

\subsection{Range of applicability}
Our method is limited in its range of applicability due to the use of a reduced speed of light. In the streaming limit, $\hat c$ may be chosen so that it is larger than the fastest radiation-driven fronts (e.g., ionization fronts, some Marshak waves) in order to maintain the correct dynamics \citep{Gnedin_2001}. In the diffusion limit, $\hat c$ may be chosen so that it is larger than the effective diffusion speed $\sim c / \tau$. In this regime, we see that energy non-conservation may cause the equilibrium temperature in a closed box to differ from the physically correct equilibrium temperature (\autoref{section:equilibrium}), but in most astrophysical RHD applications, the boundary conditions are not those of a closed system, and in this case our method recovers correct solutions without difficulty (e.g., \autoref{section:radshock}). However, as emphasized by \cite{Skinner_2013}, there is no constant choice of $\hat c$ that enables one to preserve the ordering of the hydrodynamic signal speed $c_s + |\vc{v}|$, the dynamic diffusion speed $\sim |\vc{v}| + c / \tau$, and the reduced speed of light $\hat c$ such that $v_{\text{hydro}} \ll v_{\text{diffusion}} \ll \hat c$. Extensions to the reduced speed of light method are possible that enable qualitatively correct behavior in a larger parameter space, but we leave their implementation to future work (Wibking, et al., in prep.).

\subsection{Future extensions}
There are several ways in which our code may be extended to include more physics or more accurate radiation transport. The easiest additional radiative process to include is monoenergetic, isotropic scattering, with a straightforward extension for moment methods (e.g., \citealt{Jiang12a}). Also relatively straightforward would be an extension to include coarse frequency dependence of the radiation via a multigroup extension of our radiation-matter coupling implicit solver. Even with a relatively small number of energy groups, many additional applications would be possible, including observational comparisons.

In order to improve the accuracy of the solution, one might also eschew local closures entirely and substitute a non-local closure for the Eddington tensor based on solution of the discrete ordinates ($S_N$) equations (e.g. \citealt{Davis_2012,Jiang12a}), or using our moment method as a nonlinear preconditioner to accelerate the convergence of the thermal emission term in the $S_N$ equations themselves \citep{Park_2012}. The latter is an attractive option especially when used in combination with photon-conserving spatial discretizations of the $S_N$ equations \citep{Adams97a,Adams_2001}.

In the near future, we plan to add support for self-gravity with \textsc{AMReX}'s geometric multigrid solver for GPUs \citep{AMReX_JOSS}, sink and star particles for star cluster simulations (e.g., \citealt{Krumholz04a, Offner09a}), and optically-thin line cooling for the interstellar medium. These additions will enable simulations of the interstellar medium, galactic winds, and star clusters, among others.

Radiation hydrodynamics codes like ours will enable the widespread use of more accurate radiation transport methods and an ever-greater dynamic range in both space and time. As we approach the era of exascale supercomputers, we see a bright future for AMR radiation hydrodynamics on GPU architectures.

\section*{Acknowledgements}
We thank Andrew Myers and Weiqun Zhang at LBNL for technical advice and assistance in using AMReX. BDW thanks Shyam Menon for discussions regarding the radiation-driven shell test.

This research was supported by the Australian Research Council through its Discovery Projects and Future Fellowship Funding Schemes, awards DP190101258 and FT180100375. This research was undertaken with the assistance of resources and services from the National Computational Infrastructure (NCI), which is supported by the Australian Government.

\emph{Software:} AMReX \citep{the_amrex_development_team_2021_5363443},
matplotlib \citep{Hunter:2007},
numpy \citep{harris2020array},
VisIt \citep{HPV:VisIt},
yt \citep{Turk11a}.

%%%%%%%%%%%%%%%%%%%%%%%%%%%%%%%%%%%%%%%%%%%%%%%%%%
\section*{Data Availability}
The complete source code of \textsc{Quokka}, including the source code for all test problems shown in this work, is hosted in this public \faGithub\href{https://github.com/BenWibking/quokka-code}{GitHub repository}.


\bibliographystyle{mnras}
\bibliography{quokka} % if your bibtex file is called example.bib

%%%%%%%%%%%%%%%%%%%%%%%%%%%%%%%%%%%%%%%%%%%%%%%%%%

%%%%%%%%%%%%%%%%% APPENDICES %%%%%%%%%%%%%%%%%%%%%

%%%%%%%%%%%%%%%%%%%%%%%%%%%%%%%%%%%%%%%%%%%%%%%%%%

% Don't change these lines
\bsp	% typesetting comment
\label{lastpage}
\end{document}